\documentclass[fleqn,usenatbib]{mnras}

\usepackage{newtxtext,newtxmath}
\usepackage[T1]{fontenc}

\DeclareRobustCommand{\VAN}[3]{#2}
\let\VANthebibliography\thebibliography
\def\thebibliography{\DeclareRobustCommand{\VAN}[3]{##3}\VANthebibliography}

\usepackage{graphicx}	% Including figure files
\usepackage{amsmath}	% Advanced maths commands

\usepackage{pdflscape}

\usepackage{caption}
\usepackage{subcaption}
\newcommand{\vect}[1]{\mathbf{#1}}

\title[Multiwavelength Polarization of HL Tau]{
Panchromatic (Sub)millimeter Polarization Observations of HL Tau Unveil Aligned Scattering Grains
}

\author[Z.-Y. D. Lin et al.]{
Zhe-Yu Daniel Lin,$^{1}$\thanks{E-mail: zdl3gk@virginia.edu}
Zhi-Yun Li,$^{1}$
Ian W. Stephens,$^{2}$
Manuel Fern\'andez-L\'opez,$^{3}$
Carlos Carrasco-Gonz\'alez,$^{4}$
\newauthor
Claire J. Chandler,$^{5}$
Alice Pasetto,$^{4}$
Leslie W. Looney,$^{6,7}$
Haifeng Yang,$^{8}$
Rachel E. Harrison,$^{9}$
\newauthor
Sarah I. Sadavoy,$^{10}$
Thomas Henning,$^{11}$ 
A. Meredith Hughes,$^{12}$ 
Akimasa Kataoka,$^{13,14}$
Woojin Kwon,$^{15,16}$ 
\newauthor
Takayuki Muto,$^{17,18,19}$
Dominique Segura-Cox$^{20}$\thanks{NSF Astronomy and Astrophysics Postdoctoral Fellow}
\\
$^{1}$Department of Astronomy, University of Virginia, 530 McCormick Rd., Charlottesville 22904, Virginia, USA\\
$^{2}$Department of Earth, Environment, and Physics, Worcester State University, Worcester, Massachusetts 01602, USA\\
$^{3}$Instituto Argentino de Radioastronom{\'i}a, CCT-La Plata (CONICET), C.C.5, 1894, Villa Elisa, Argentina\\
$^{4}$Instituto de Radioastronom\'ia y Astrof\'isica (IRyA-UNAM), Morelia, Mexico\\
$^{5}$National Radio Astronomy Observatory, PO Box O, Socorro, NM 87801, USA\\
$^{6}$Department of Astronomy, University of Illinois, 1002 W Green St., Urbana, Illinois 61801, USA\\
$^{7}$National Radio Astronomy Observatory, 520 Edgemont Rd., Charlottesville, Virginia 22903, USA\\
$^{8}$Kavli Institute for Astronomy and Astrophysics, Peking University, Yi He Yuan Lu 5, Haidian Qu, Beijing 100871, People’s Republic of China\\
$^{9}$School of Physics and Astronomy, Monash University, Clayton VIC 3800, Australia\\
$^{10}$Department of Physics, Engineering Physics and Astronomy, Queen’s University, Kingston, ON, K7L 3N6, Canada\\
$^{11}$Max Planck Institute for Astronomy, K{\"o}nigstuhl 17, D-69117 Heidelberg, Germany\\
$^{12}$Van Vleck Observatory, Wesleyan University, 96 Foss Hill Dr, Middletown, Connecticut 06459, USA\\
$^{13}$Department of Astronomical Science, Graduate University for Advanced Studies (SOKENDAI), 2-21-1 Osawa, Mitaka, Tokyo 181-8588, Japan\\
$^{14}$National Astronomical Observatory of Japan, 2-21-1 Osawa, Mitaka, Tokyo 181-8588, Japan\\
$^{15}$Department of Earth Science Education, Seoul National University, 1 Gwanak-ro, Gwanak-gu, Seoul 08826, Republic of Korea\\
$^{16}$SNU Astronomy Research Center, Seoul National University, 1 Gwanak-ro, Gwanak-gu, Seoul 08826, Republic of Korea\\
$^{17}$Division of Liberal Arts, Kogakuin University, 1-24-2 Nishi-Shinjyuku, Shinjyuku-ku, Tokyo 163-8677, Japan\\
$^{18}$Leiden Observatory, Leiden University, P.O. Box 9513, NL-2300 RA Leiden, The Netherlands\\
$^{19}$Department of Earth and Planetary Sciences, Tokyo Institute of Technology, 2-12-1 Oh-okayama, Meguro-ku, Tokyo 152-8551, Japan\\
$^{20}$Department of Astronomy, The University of Texas at Austin, 2515 Speedway, Austin, Texas 78712, USA
}

\date{Accepted XXX. Received YYY; in original form ZZZ}

\pubyear{2015}

\begin{document}
\label{firstpage}
\pagerange{\pageref{firstpage}--\pageref{lastpage}}
\maketitle

\begin{abstract}
Polarization is a unique tool to study the properties of dust grains of protoplanetary disks and detail the initial conditions of planet formation. Polarization around HL Tau was previously imaged using the Atacama Large Millimeter/submillimeter Array (ALMA) at Bands 3 (3.1~mm), 6 (1.3~mm), and 7 (0.87~mm), showing that the polarization orientation changes across wavelength $\lambda$. The polarization morphology at Band 7 is predominantly parallel to the disk minor axis but appears azimuthally oriented at Band 3, with the morphology at Band 6 in between the two. We present new $\sim 0.2\arcsec$ (29~au) polarization observations at Q-Band (7.0~mm) using the Karl G. Jansky Very Large Array (VLA) and at Bands 4 (2.1~mm), 5 (1.5~mm), and 7 using ALMA, consolidating HL Tau's position as the protoplanetary disk with the most complete wavelength coverage in dust polarization. The polarization patterns at Bands 4 and 5 continue to follow the morphological transition with wavelength previously identified in Bands 3, 6, and 7. Based on the azimuthal variation, we decompose the polarization into contributions from scattering ($s$) and thermal emission ($t$). We find that $s$ decreases slowly with increasing $\lambda$, and $t$ increases more rapidly with $\lambda$ which are expected from optical depth effects of toroidally aligned, scattering prolate grains. The relatively weak $\lambda$ dependence of $s$ is consistent with large, porous grains. The sparse polarization detections from the Q-band image are also consistent with toroidally aligned prolate grains. 

\end{abstract}

\begin{keywords}
polarization -- protoplanetary discs -- ISM: individual objects: HL Tau
\end{keywords}

%%%%%%%%%%%%%%%%%%%%%%%%%%%%%%%%%%%%%%%%%%%%%%%%%%

%%%%%%%%%%%%%%%%% BODY OF PAPER %%%%%%%%%%%%%%%%%%

\section{Introduction}

Studying the dust properties of protoplanetary disks is crucial for understanding the origins of planets, because dust grains serve as the building blocks of planet formation \citep[e.g.][]{Beckwith2000prpl.conf..533B, Johansen2014prpl.conf..547J, Morbidelli2016JGRE..121.1962M}. Polarization at millimeter wavelengths has emerged as a unique and powerful tool for studying the properties of dust grains and their initial conditions in disks \cite[e.g.][]{Andersson2015ARA&A..53..501A, Kataoka2015ApJ...809...78K}. With the advent of the Atacama Large Millimeter/submillimeter Array (ALMA), the field of (sub)millimeter-wavelength disk polarization has witnessed a revolution, thanks to the unprecedented sensitivity and spatial resolution \citep[e.g.][]{Kataoka2016ApJ...831L..12K, Stephens2017, Alves2018A&A...616A..56A, Lee2018ApJ...854...56L, Girart2018ApJ...856L..27G,  Bacciotti2018ApJ...865L..12B, Dent2019MNRAS.482L..29D, Takahashi2019ApJ...872...70T, Harrison2019ApJ...877L...2H, Sadavoy2019ApJS..245....2S, Harrison2021ApJ...908..141H, Ohashi2020ApJ...900...81O, Stephens2020ApJ...901...71S, Lee2021ApJ...910...75L, Aso2021ApJ...920...71A, Tang2023ApJ...947L...5T}. 

A common process to produce disk polarization is through dust scattering. Grains can efficiently scatter thermal radiation from other grains when the sizes of grains become comparable to the observing wavelength \citep{Bohren1983asls.book.....B, Kataoka2015ApJ...809...78K}. This mechanism produces a distinctive pattern in an inclined disk where the polarization direction is parallel to the disk minor axis \citep{Yang2016MNRAS.456.2794Y, Kataoka2016ApJ...820...54K}. Most sources with resolved disk-scale polarization observations show this pattern \citep[e.g.][]{Stephens2014Natur.514..597S, Stephens2017, Hull2018ApJ...860...82H, Takahashi2019ApJ...872...70T} and the measurements of the spectral index of Stokes~$I$ support the dust scattering interpretation \citep[e.g.][]{Zhu2019ApJ...877L..18Z, Liu2019ApJ...877L..22L, CarrascoGonzalez2019ApJ...883...71C, Lin2020MNRAS.496..169L}. 

Another process to produce polarization is through polarized thermal emission of aligned, elongated grains. There are several proposed mechanisms to align grains, including radiative alignment torques (RAT; \citealt{Dolginov1976Ap&SS..43..291D, Draine1997ApJ...480..633D}), mechanical alignment torques (MET; \citealt{Gold1952MNRAS.112..215G, Lazarian2007ApJ...669L..77L, Hoang2018ApJ...852..129H}), or paramagnetic alignment, which can align grains either to the magnetic field, radiation field, or the gas flow depending on the details of each mechanism \citep[e.g.][]{Andersson2015ARA&A..53..501A}. While grains are likely aligned to the magnetic field in the diffuse ISM and protostellar envelopes through RAT \citep[e.g.][]{Legouellec2020A&A...644A..11L, Valdivia2022A&A...668A..83V}, it is unclear which mechanism can align grains in protoplanetary disks. Nevertheless, one can infer the presence of aligned grains through a consistent polarization pattern across wavelengths \citep{Cox2015ApJ...814L..28C, Alves2018A&A...616A..56A} or through a 90$^{\circ}$ flip due to dichroic extinction \citep{Ko2020ApJ...889..172K, Lin2020MNRAS.493.4868L, Liu2021ApJ...914...25L}.

Interestingly, in some disks, polarization measurements exhibit polarization consistent with dust scattering at shorter wavelengths, but the polarization becomes azimuthally oriented at longer wavelengths \citep[e.g.][]{Stephens2017, Harrison2019ApJ...877L...2H, Mori2019ApJ...883...16M, Harrison2021ApJ...908..141H}. The difference in the polarization patterns is not expected from scattering or aligned grains alone \citep{Yang2016MNRAS.460.4109Y, Stephens2017, Yang2017MNRAS.472..373Y, Mori2021ApJ...908..153M}. The best-studied case, thus far, that exhibits the transition in disk-scale polarization morphology with wavelength is HL Tau, a Class I/II protostar. At Band~3 the polarization is azimuthally oriented with $\sim 2\%$ polarization \citep{Kataoka2017ApJ...844L...5K}. At Band~7, the polarization becomes unidirectional and parallel to the disk minor axis with $\sim 0.8\%$ polarization \citep{Stephens2014Natur.514..597S, Stephens2017}. Intriguingly, the Band~6 image has polarization directions that are in between the two extremes \citep{Stephens2017}.

Studies have shown that the azimuthally oriented polarization at Band~3 seen in HL Tau is better explained by toroidally aligned prolate grains than radially aligned oblate grains based on the azimuthal variation of polarization \citep{Kataoka2017ApJ...844L...5K, Yang2019MNRAS.483.2371Y, Mori2021ApJ...908..153M}. By self-consistently solving radiation transfer equation including the thermal polarization and scattering of aligned grains, \cite{Lin2022} demonstrated that the transition in polarization morphology could be attributed to an increase of optical depth towards shorter wavelengths that causes scattering polarization to dominate over the polarization from the underlying thermal polarization of aligned grains. The optical depth interpretation also naturally explains the Band~6 image that appears in between the two extreme morphology if the optical depth is largely in between that at Bands~3 and 7. To further test if toroidally aligned prolate grains with varying optical depth can explain the polarization transition, we need additional resolved polarization observations at different wavelengths.

HL Tau is located in the L1551 dark cloud of the Taurus-Auriga molecular cloud complex \citep{Kenyon2008hsf1.book..405K}. The conventional adopted distance for the cloud complex is $140$~pc \citep{Kenyon1994AJ....108..251K}, but recent advancements in distance measurement have revealed a significant line of sight depth \citep{Loinard2013IAUS..289...36L}. Studies utilizing Gaia data have reported distances of 145~pc \citep{Luhman2018AJ....156..271L} and $146 \pm 0.6$~pc \citep{Roccatagliata2020A&A...638A..85R}. Additionally, the Very Long Baseline Array yielded a distance of $147.3 \pm 0.5$~pc \citep{Galli2018ApJ...859...33G}. We adopt a distance of 147.3 pc for HL Tau for consistency with the recent high angular resolution study \citep{CarrascoGonzalez2019ApJ...883...71C}.

In this paper, we present new polarization observations at Bands 4 and 5 using ALMA and Q-Band using the Very Large Array (VLA) to investigate whether the observed transition in polarization extends to other wavelengths. We also present a new ALMA Band 7 polarization image with improved angular resolution and reprocessed previous ALMA Bands 3 and 6 data gathering a final set of images with comparable angular resolution. By obtaining multiwavelength polarization images, we aim to confirm the presence of the transition and test predictions from optical depth effects \citep{Lin2022}. The paper is organized as follows: In Section~\ref{sec:observations}, we provide a brief overview of the observations and the data calibration procedure. Section~\ref{sec:results} presents our results, showcasing the polarization properties of HL Tau at different wavelengths, and we analyze the polarization across wavelengths in Section~\ref{sec:analysis}. We discuss the implications of our results in Section~\ref{sec:discussion} and summarize in Section~\ref{sec:conclusions}.

\section{Observations} \label{sec:observations}
To date, HL Tau has been observed by ALMA at Bands 3 (3.09~mm), 4 (2.07~mm), 5 (1.48~mm), 6 (1.29~mm), and 7 (0.87~mm) and by the VLA at Q-band (6.97~mm). Bands 3 (project code: 2016.1.00115.S; PI: Akimasa Kataoka) and 6 (project code: 2016.1.00162.S; PI: Ian Stephens) data were first presented in \cite{Kataoka2017ApJ...844L...5K} and \cite{Stephens2017}, respectively, but we reimaged the measurement sets after self-calibration. While Band~7 was originally presented in \cite{Stephens2017}, we used deeper and higher resolution data from Stephens et al., (in press) (project code: 2019.1.01051.S; PI: Ian Stephens). Table~\ref{tab:calibrators} is the observation log which lists the relevant observation settings, including the bandpass, amplitude, phase, and polarization calibrators. We used the Common Astronomy Software Applications (CASA) package for all calibration and imaging on the ALMA and VLA data \citep{McMullin2007ASPC..376..127M}. 

\subsection{ALMA Observations}
For all the ALMA data presented in this paper, including archival and new data, we self-calibrated and imaged the data for all 5 bands so that they would all be imaged in a consistent manner. Prior to self-calibration, we re-ran the data through ALMA's calibration pipeline using the ALMA-supplied calibration scripts. These scripts do the standard calibration, which includes bandpass, phase, polarization, and flux calibration.

To run the calibration pipeline we used CASA version 4.7.38335 for Band~3, while for Bands~4,~5, and~7 we used version 6.2.1.7. The calibrated Band~6 dataset was provided by the ALMA Helpdesk staff. Line removal, self-calibration, and imaging were performed using the CASA version 6.2.1.7 for all the bands. Every dataset of each band consists of four 2~GHz spectral windows with 64 channels. The total effective bandwidth of each dataset is approximately 7.5~GHz. However, we identified some prominent molecular lines that we removed when making the continuum images. While we did not find significant line emission in the Band~3 data, we identified the SO(3,4 -- 2,3) line at $\nu_{\text{rest}}=138.179$~GHz in Band~4. We also identified: CS(4 -- 3) at $\nu_{\text{rest}}=195.954$~GHz in Band~5; CH$_3$OH(20,-2,19 -- 19,-3,17) and H$_2$CO(3,1,2 -- 2,1,1) at $\nu_{\text{rest}}=224.700$~GHz and $\nu_{\text{rest}}=225.698$~GHz, respectively, in Band~6; C$^{17}$O(3 -- 2), SO$_2$(18,4,14 -- 18,3,15), SO(3,3 -- 2,3), CH$_3$OH(9,5,5 -- 10,4,6), and SO$_2$(5,3 -- 4,2) at $\nu_{\text{rest}}=337.061$~GHz, $\nu_{\text{rest}}=338.306$~GHz, $\nu_{\text{rest}}=339.342$~GHz,  $\nu_{\text{rest}}=351.236$~GHz, and $\nu_{\text{rest}}=351.257$~GHz, respectively, in Band~7.

We used a similar standard self-calibration procedure for every band dataset. We use \textit{tclean} for imaging and use the Briggs robust parameter of 0.5 for each wavelength. The data from every band went through three rounds of phase-only self-calibration, with solution intervals \textit{infinity}, 30.5\,s, and 10.4\,s. Final deep cleaning of the four Stokes parameters using a cleaning mask covering the HL~Tau disk area led to signal-to-noise ratios of $\sim 1200$, 890, 1200, 1100, and 1300 from Bands 3 to 7, respectively. Table~\ref{tab:basic_image_parameters} lists the resulting synthesized beam sizes.

\subsection{VLA Q-band Observations}

We observed HL Tau with the VLA in its B configuration during three semesters (Legacy project code: 19A-388). We completed eight observation epochs between May 2019 to September 2021 (2 in 2019, 5 in 2020, and 1 in 2021). We used the usual continuum frequency setup covering a frequency range 39-47 Hz, and full polarization mode. In each epoch, the total observing time was 5 hours with 2.5 hours on target. In all epochs, the flux calibrator was 3C147, bandpass calibrator was 3C84, and gain calibrator (observed every 45s) was J0431+1731. For the calibration of the data we used CASA and a modified version of the NRAO Pipeline which includes polarization calibration after the usual gain calibration. For the calibration of the polarization angle, we used the known polarization parameters for 3C147, i.e., a polarization angle of 86$^\circ$ and a polarization degree of 5.2\% (Perley \& Butler 2013). We assumed these parameters to be constant across the 8 GHz bandwidth of the Q band observations. For the calibration of the leakage terms, we used the gain calibrator, J0431+1731, which was always observed for a wide range of parallactic angles. We assumed an unknown polarization for this calibrator and solved for it. We checked the consistency of polarization parameters of the leakage calibrator at each epoch, and discarded one epoch due to very different values of the polarization angle and polarization degree. After initial calibration, we corrected for small shifts in the position of the source in each epoch. The final, aligned, and concatenated dataset contains 17.5 hours on target. The final images were made using \emph{tclean} and a natural weighting. The signal-to-noise ratio of the peak $I$ is 210. The resulting synthesized beam size is $0.156 \arcsec \times 0.143 \arcsec$  (Table~\ref{tab:basic_image_parameters}).

\subsection{Construction of Polarization Images}
The basic statistics of the images are recorded in Table~\ref{tab:basic_image_parameters}. The noise levels for each Stokes parameter, $I$, $Q$, $U$, and $V$, are denoted as $\sigma_{I}$, $\sigma_{Q}$, $\sigma_{U}$, and $\sigma_{V}$, respectively. 
$F_{\nu}$ is the flux density of Stokes~$I$ where we use emission above 3~$\sigma_{I}$. We assume a 10$\%$ absolute calibration uncertainty based on the VLA and the ALMA technical handbooks, but we ignore it for the rest of the paper.

In the ideal limit without noise, the linear polarized intensity is directly related to Stokes~$Q$ and $U$ through:
\begin{align} \label{eq:P_ideal_limit}
    P_{m} \equiv \sqrt{Q^{2} + U^{2}}. 
\end{align}
However, when including noise, Eq.~(\ref{eq:P_ideal_limit}) results in a positive bias, because the Stokes~$Q$ and $U$ can be positive or negative while the linear polarized intensity is always positive.

Following \cite{Vaillancourt2006PASP..118.1340V} and \cite{Hull2015JAI.....450005H}, we debias the linear polarized intensity by considering the probability density function (PDF):
\begin{align} \label{eq:pdf_of_true_P}
    \text{PDF}(P | P_{m}, \sigma_{P}) = \frac{P}{ \sigma_{P}^{2} } I_{0} \bigg(\frac{ P P_{m} }{ \sigma_{P}^{2} } \bigg) \exp{ \bigg[ - \frac{ ( P_{m}^{2} + P^{2} )  }{ 2 \sigma_{P}^{2} }  \bigg] }
\end{align}
which describes the probability of the true linear polarized intensity $P$ given a measured $P_{m}$ and noise level $\sigma_{P}$. $I_{0}$ is the zeroth-order modified Bessel function of the first kind. $\sigma_{P}$ comes from $\sigma_{Q}$ and $\sigma_{U}$ which are usually comparable, but we define the noise level of the linear polarized intensity through
\begin{align} \label{eq:lpi_rms}
    \sigma_{P} = \sqrt{(\sigma_{Q}^{2} + \sigma_{U}^{2}) / 2} 
\end{align}
as an explicit way to account for any slight difference. Thus, we obtain $P$ by finding the maximum of Eq.~(\ref{eq:pdf_of_true_P}). For high signal-to-noise detections ($P_{m} \geq 5\sigma_{P}$), a simple approximation exists: 
\begin{align} \label{eq:P_high_SNR}
    P = \sqrt{ Q^{2} + U^{2} -  \sigma_{P}^{2} }, 
\end{align}
but we use Eq.~(\ref{eq:pdf_of_true_P}) for $P_{m}<5\sigma_{P}$. 

The sign of the Stokes parameters follows the IAU convention \citep{Contopoulos1974tiau.book.....C, Hamaker1996A&AS..117..137H, Hamaker1996A&AS..117..161H}. 
The polarization angle is defined by 
\begin{align} \label{eq:polarization_angle}
    \chi \equiv \frac{1}{2} \arctan \bigg(\frac{U}{Q} \bigg)
\end{align}
and goes East-of-North. We only consider the E-vectors, whose angles are defined by Eq.~(\ref{eq:polarization_angle}) and not the B-vectors (rotated by 90$^{\circ}$) that are conventionally used to trace the magnetic field assuming aligned oblate grains. The uncertainty of $\chi$ is 
\begin{align}
    \sigma_{\chi} = \frac{1}{2} \frac{ \sigma_{P} }{ P } 
\end{align}
\citep{Hull2015JAI.....450005H}. 

We further define several convenient quantities. The linear polarization fraction is
\begin{align}
    p &\equiv \frac{P}{I}. 
\end{align}
In addition, the Stokes~$Q$ and $U$ normalized by Stokes~$I$ are $q \equiv Q / I$ and $u \equiv U/I$, where we use lowercase to represent quantities of polarized intensity normalized by Stokes~$I$. 

The uncertainty of $p$ is
\begin{align} \label{eq:error_lpol}
    \sigma_{\text{pf}} &= \frac{P}{I} \sqrt{ \bigg( \frac{ \sigma_{P} }{ P } \bigg)^{2} + \bigg(\frac{\sigma_{I}}{I} \bigg)^{2}}
\end{align}
which is estimated through error propagation. We note that the ALMA technical handbook gives a minimum detectable degree of polarization, which is defined as three times the systematic calibration uncertainty, of $0.1\%$ for compact sources within the inner third of the primary beam. Thus, we use the error of $0.033\%$ whenever the error from Eq.~(\ref{eq:error_lpol}) is less than this value for data from ALMA. The uncertainties of $q$ and $u$ are likewise estimated through error propagation.

\begin{landscape}
\begin{table}
    \centering
    \begin{tabular}{c c c c c c c c c c c c c}
        \hline
        Band & EBs & UTC Date & T$_{\text{on-sou}}$ & Config & N$_{\text{ant}}$ & Baselines & Bandpass & Amplitude & Phase & Polarization & Project Code \\
         & & & (hours) & & & (m) &  & & \\
        (1) & (2) & (3) & (4) & (5) & (6) & (7) & (8) & (9) & (10) & (11) & (12) \\
        \hline 
        Q & 8 & 2019 May - 2021 Sept & 17.5 & B & 27 &  133-11126 &  3C84 & 3C147 & J0431+1731 & 3C147, J0431+1731 & 19A-388\\
        3 & 4 & 2016 Oct 12 & 2.3 & C40-6 & 41 & 19-3144 & J0510+1800 & J0423-0120 & J0431+1731 & J0510+1800 & 2016.1.00115.S \\
        4 & 2 & 2021 Jul 7 & 1.2 & C43-6/7 & 38 & 41-3396 & J0510+1800 & J0510+1800 & J0431+1731 & J0423-0120 & 2019.1.00134.S \\
        5 & 2 & 2021 Jun 14 & 1.1 & C43-6 & 41 & 16-2517 & J0238+1636 & J0238+1636 & J0431+1731 & J0423-0120 & 2019.1.00134.S\\
         & 1 & 2021 Jul 6 & 0.5 & C43-6/7 & 41 & 41-3638 & J0510+1800 & J0510+1800 & J0431+1731 & J0423-0120 & 2019.1.00134.S\\
        6 & 3 & 2017 Jul 12 & 1.1 & C40-5 & 42 & 17-2647 & J0510 +1800 & J0510+1800 & J0431+1731 & J0522–3627 & 2016.1.00162.S \\
        7 & 2 & 2021 Jun 30 & 1.2 & C43-6 & 41 & 15-2114 & J0510+1800 & J0510+1800 & J0431+1731 & J0423-0120 & 2019.1.01051.S \\
        \hline 
    \end{tabular}
    \caption{
        Column 1: Name of the band. 
        Column 2: Number of Execution Blocks per project. 
        Column 3: Observation start date in UTC. 
        Column 4: Time on source in hours. 
        Column 5: Antenna configuration. 
        Column 6: Number of antenna used. 
        Column 7: Range of baselines in meters. 
        Columns 8, 9, 10 and 11: Quasars used for bandpass, flux, phase, and polarization calibration.        
        Column 12: The associated project code.
        %PWV: 1.54, 0.26, 0.60, 0.75, 1.52, 0.46 in mm for ALMA
    }
    \label{tab:calibrators}
\end{table}
\end{landscape}

\begin{landscape}
    \begin{table}
        \centering
        \begin{tabular}{c c c c c c c c c c c c }
            \hline
            Band & $\lambda$ & Beam Major & Beam Minor & Beam PA & $\sigma_{I}$ & $\sigma_{Q}$ & $\sigma_{U}$ & Peak $I$ & Peak $P$ & Median $p$ & $F_{\nu}$ \\
            & mm & $\arcsec$ & $\arcsec$ & $^{\circ}$ & $\mu$Jy/beam & $\mu$Jy/beam & $\mu$Jy/beam & mJy/beam & $\mu$Jy/beam & $\%$ & mJy \\
            (1) & (2) & (3) & (4) & (5) & (6) & (7) & (8) & (9) & (10) & (11) & (12)  \\
            \hline
            Q & 6.97 & 0.16 & 0.14 & 45 & 4.9 & 3.9 & 4.0 & 1.010 & 17 & 6.7 & 4.94 \\
            3 & 3.08 & 0.43 & 0.29 & -13 & 21 & 7.0 & 7.0 & 25.45 & 144 & 1.8 & 75.0 \\
            4 & 2.07 & 0.21 & 0.19 & -28 & 31 & 7.7 & 7.7 & 27.53 & 186 & 1.6 & 215 \\
            5 & 1.48 & 0.19 & 0.16 & -76 & 40 & 12 & 12 & 47.16 & 360 & 1.1 & 525 \\
            6 & 1.29 & 0.27 & 0.16 & -46 & 72 & 15 & 15 &  81.78 & 590 & 0.88 & 710 \\
            7 & 0.872 & 0.20 & 0.13 & -81 & 92 & 25 & 24 & 121.16 & 740 & 0.87 & 1880 \\
            \hline
        \end{tabular}
        \caption{
            Basic statistics of each image at different bands. Column 1: Name of the wavelength band. Column 2: Representative wavelength of the continuum. Columns 3 and 4: the FWHM along the major and minor axes of the beam. Column 5: Position angle (East-of-North) of the beam. Column 6, 7, and 8: The noise levels for Stokes~$IQU$, respectively. Column 9: Peak of the Stokes~$I$ image. Column 10: Peak of the $P$ image. Column 11: Median of the $p$ image for regions with detection. Column 12: $F_{\nu}$ is the flux density integrated from emission above 3~$\sigma_{I}$.
        }
        \label{tab:basic_image_parameters}
    \end{table}
\end{landscape}

\section{Results} \label{sec:results}

\subsection{Polarization Morphology}
% also talk about the peak flux, flux density, 

% check if the fluxes are consistent with past papers 

\begin{figure*}
    \centering
    \includegraphics[width=\textwidth]{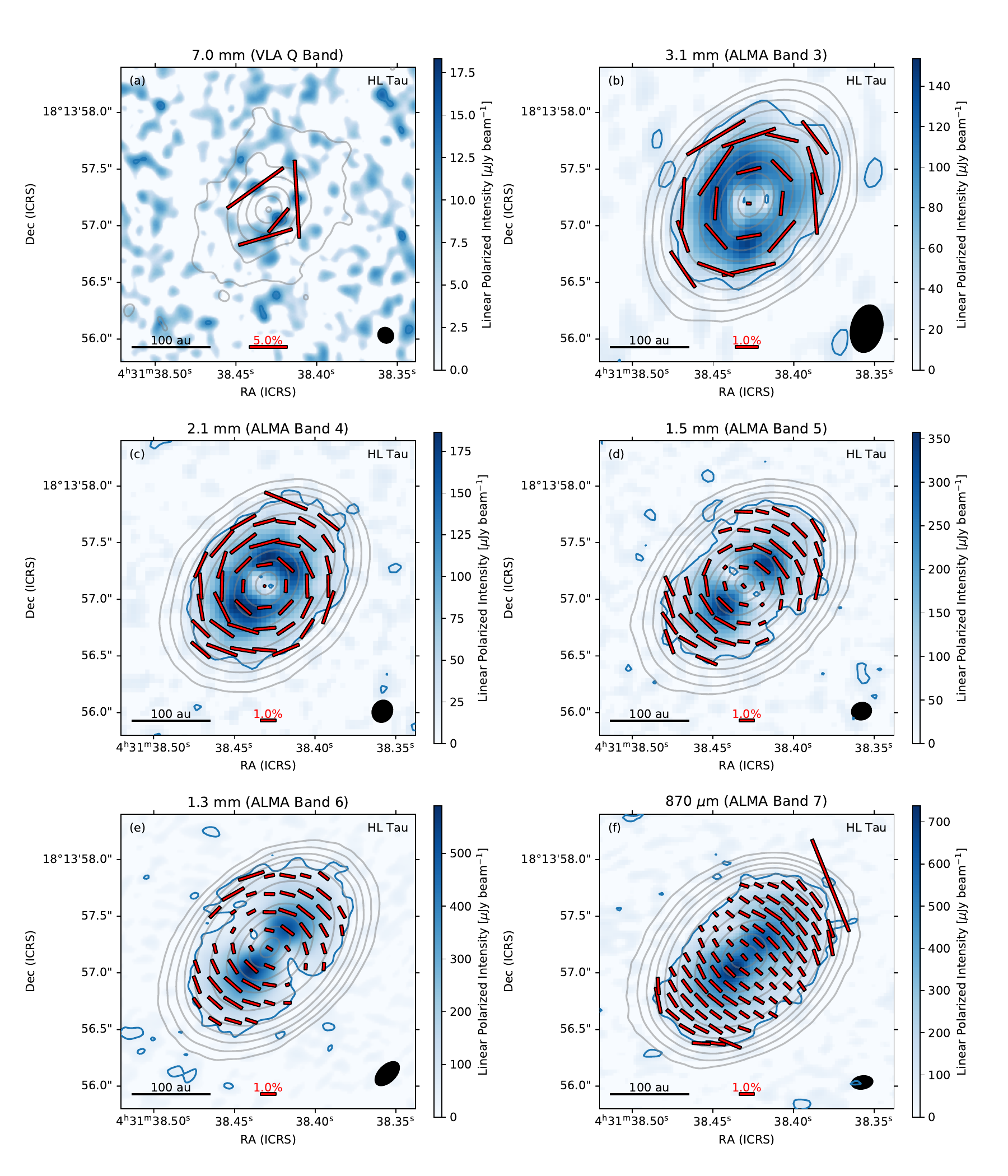}
    \caption{
        Panels a to f show the polarimetric data from the VLA Band Q and ALMA Bands 3, 4, 5, 6, and 7, respectively. In each panel, the color map represents the linear polarized intensity in $\mu$Jy beam$^{-1}$. The blue contour traces the $3\sigma_{P}$ level, while grey contours show the Stokes~$I$ in steps of 3, 10, 25, 50, 100, 200, 325, 500, 750, and 1000 $\sigma_{I}$. The direction of the red line segments represents the polarization angle, while the length of the line segments is proportional to the linear polarization fraction. Each line segment samples the image in step sizes equal to the FWHM of the minor axis of the beam. The length of $1\%$ polarization is shown in the center bottom. The black bar to the bottom left shows the 100~au scale. The black ellipse to the bottom right represents the synthesized beam.
    }
    \label{fig:complete_wcs}
\end{figure*}

Fig.~\ref{fig:complete_wcs} shows the polarization images across all six bands. There exists a consistent transition in the polarization morphology across the spectrum. Starting from the longest wavelength with Fig.~\ref{fig:complete_wcs}a, the VLA Q-Band only marginally detected a few vectors (E-vectors). Although there are a few regions with $P$ above $3\sigma_{P}$ in the image, we only consider polarization detections where Stokes~$I$ is also detected above $3\sigma_{I}$. 
The vector closest to the center is $\sim 4\%$ and appears parallel to the disk major axis. The other vectors are $\sim 10\%$ and are oriented azimuthally around the center.

The image at $\lambda$=3.1~mm (Band 3) shows an azimuthal distribution of $P$ around a center of low $P$ with two null points to the East and West of the center. The polarization direction ($E$-vectors) is oriented azimuthally around the center in that the polarization along the major axis is parallel to the disk minor axis and that along the minor axis is parallel to the disk major axis. In addition, the polarization fraction $p$ is larger at larger radii. These characteristics are qualitatively consistent with \cite{Kataoka2017ApJ...844L...5K} and \cite{Stephens2017} where the data originally appeared. The resolution of $\sim 0.35\arcsec$ in this work is similar to that in \cite{Kataoka2017ApJ...844L...5K} which also used robust=0.5 and is slightly better than the resolution of $\sim 0.46\arcsec$ in \cite{Stephens2017} which used robust=1.0.

The image at 2.1~mm (Band~4) appears similar to the Band 3 image in that $P$ is azimuthally distributed around the center and the polarization vectors are also directed azimuthally. The main difference is that $P$ is slightly separated into two lobes along the major axis of the disk, whereas $P$ at Band 3 appears relatively more uniform. 

The 1.5~mm (Band~5) image shows a more obvious change in the distribution of $P$ and in the polarization angle. $P$ is clearly stronger along the major axis than along the minor axis. The two lobes along the major axis are more obvious and a weak link at the center emerged, forming a ``dumbbell'' shape. Along the disk minor axis, we detect polarization in the northeast (beyond the null point) with polarization parallel to the disk major axis, while $P$ at the corresponding location in the southwest is less well detected.

At 1.3~mm (Band~6), the image also shows a stronger $P$ along the major axis than along the minor axis, with a prominent dumbbell shape similar to that at Band 5. Also, the polarization vectors are clearly no longer directed azimuthally like at 3.1~mm. Instead, the vectors around the northeast edge and the southwest edge appear tilted towards the disk minor axis. The Band~6 image in this work is qualitatively similar with \cite{Stephens2017} where the data originally appeared, but differs in angular resolution in that the previous work used robust=1.0. We also better detect $P$ in the northeastern part of the disk minor axis resulting in a reduced null point. 

At 870~$\mu$m (Band~7), $P$ is distributed across the disk without any null points and the polarization is mostly parallel to the disk minor axis with slight deviations that resemble the elliptical pattern at longer wavelengths. The resolution is better than the one in \cite{Stephens2017} ($\sim 0.39 \arcsec$). The high polarization vectors in the southwest location in \cite{Stephens2017} do not appear in the new image which could suggest a spurious detection. The uniform polarization morphology across the disk is similar to the polarization expected from scattering in an inclined disk \citep{Yang2016MNRAS.456.2794Y}.

%Across all wavelengths, the morphology of Stokes~$I$ does not change much, but the morphology of polarization changes drastically.

\subsection{Individual Polarization Quantities}

\begin{figure*}
    \centering
    \includegraphics[width=\textwidth,trim={0cm 0cm 0.5cm 0cm},clip]{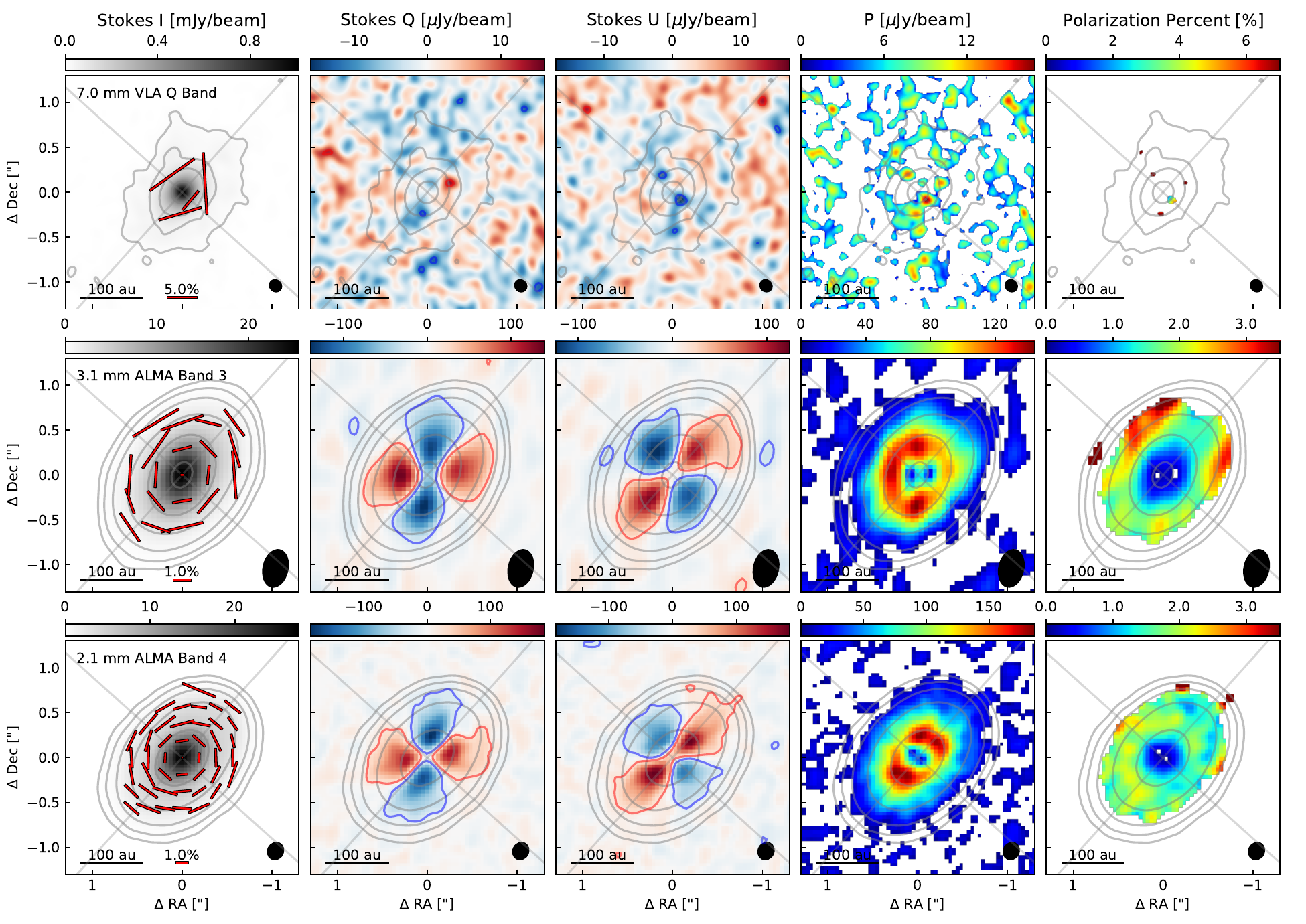}
    \caption{
        The Stokes $IQU$, linear polarized intensity $P$, and linear polarization percent $p$ images from the left to right columns. The wavelengths, from the top to the bottom row, are VLA Q-Band and ALMA Bands 3, 4, 5, 6, and 7. The vertical axis of the image is the direction to the north and increases to the top. The horizontal axis is the direction to the east and increases to the left. The line segments on top of the Stokes~$I$ images represent the polarization direction and the segment length is proportional to $p$ where the scale bar is shown at the bottom. The color scales of Stokes~$QU$ are plotted such that the white corresponds to the zero level. The $-3\sigma$ and $3\sigma$ levels are marked by blue and red contours, respectively. The synthesized beam is represented as a black ellipse to the lower right of each plot. 
    }
    \label{fig:obs_stokes_1}
\end{figure*}

To allow for a more complete view of the intricate changes across wavelength, Fig.~\ref{fig:obs_stokes_1} shows all the Stokes parameters along with $P$ and $p$. For better comparison across these observations which were taken at different dates, we fit a 2D Gaussian to the Stokes~$I$ of each band and set the center of the fitted 2D Gaussian as the origin of the image. We use the CASA task \textit{imfit} for the fitting and we use $\sigma_{I}$ (Table~\ref{tab:basic_image_parameters}) as the input noise level. Table~\ref{tab:gaussian_fitting} lists the resulting best-fit value and uncertainty of the center, deconvolved major and minor FWHM, and the position angle.

%The deconvolved major axis decreases with increasing wavelength which is expected from decreasing optical depth effects.  The position angle appears to change from $142^{\circ}$ at Q-Band to $137.21^{\circ}$ at Band~$7$. The position angle of the disk major axis is $138.02^{\circ}$ based on high angular resolution images from \cite{ALMApartnership2015}. The inclination gradually changes with wavelength. 

\begin{table*}
    \centering
    \begin{tabular}{cccccc}
        \hline 
        Band & ICRS R.A. & ICRS Dec & Major & Minor & PA \\
         & (h m s) & (d m s) & (mas) & (mas) & (degrees)  \\
        (1) & (2) & (3) & (4) & (5) & (6) \\
        \hline 
        Q & 04:31:38.429 & +18:13:57.16 & 325 $\pm$ 2 & 244 $\pm$ 2 & 142 $\pm$ 1 \\
        3 & 04:31:38.428 & +18:13:57.20 & 606 $\pm$ 1 & 422.0 $\pm$ 0.8 & 140.5 $\pm$ 0.2  \\
        4 & 04:31:38.431 & +18:13:57.12 & 674.0 $\pm$ 0.9 & 467.9 $\pm$ 0.6 & 138.6 $\pm$ 0.1  \\
        5 & 04:31:38.431 & +18:13:57.12 & 765.4 $\pm$ 0.6 & 524.7 $\pm$ 0.4 & 137.83 $\pm$ 0.09  \\
        6 & 04:31:38.428 & +18:13:57.22 & 800.08 $\pm$ 0.78 & 541.6 $\pm$ 0.5 & 137.36 $\pm$ 0.09  \\
        7 & 04:31:38.430 & +18:13:57.14 & 887.91 $\pm$ 0.57 & 606.3 $\pm$ 0.4 & 137.21 $\pm$ 0.07 \\
        \hline
    \end{tabular}
    \caption{ 
        Results from fitting the image with a 2D Gaussian. Column 1: Name of the band. Columns 2 and 3: The R.A. and Dec. of the center of the 2D Gaussian. Column 4: The deconvolved FWHM along the major axis in mas. Column 5: The deconvolved FWHM along the minor axis in mas. Column 6: The position angle of the major axis of the 2D Gaussian (East-of-North).
    }
    \label{tab:gaussian_fitting}
\end{table*}

\renewcommand{\thefigure}{\arabic{figure}}
\addtocounter{figure}{-1}
\begin{figure*}
    \centering
    \includegraphics[width=\textwidth,trim={0cm 0cm 0.5cm 0.cm},clip]{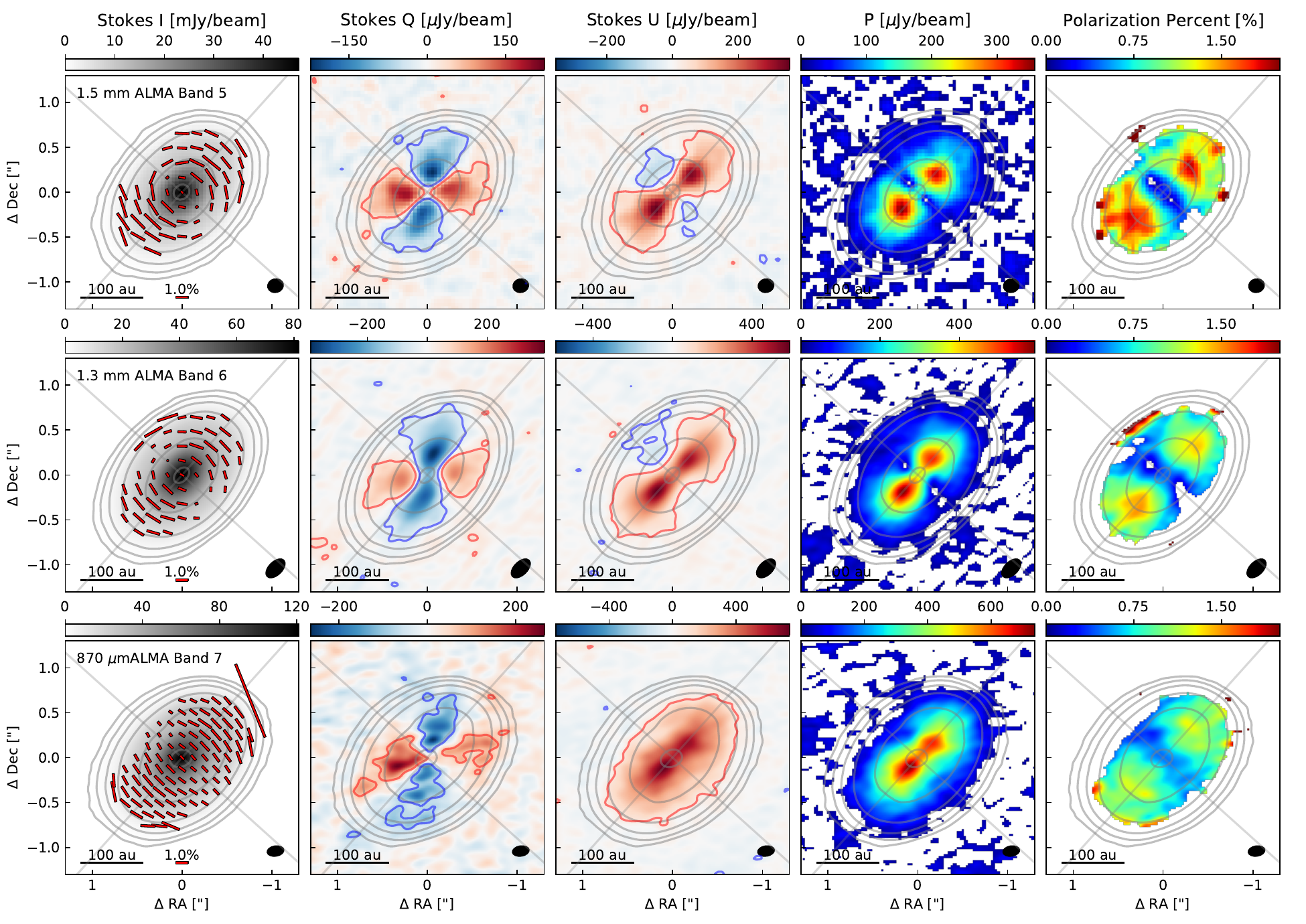}
    \caption{
          continued
    }
    \label{fig:obs_stokes_2}
\end{figure*}
\renewcommand{\thefigure}{\arabic{figure}}

The morphology of Stokes~$Q$ (Fig.~\ref{fig:obs_stokes_1}, second column) does not change much across the ALMA bands. The Stokes~$Q$ map is negative along the north-south line and positive along the east-west line with respect to the center. Recall that $+Q$ means a polarization direction that is parallel to the Dec. axis and $-Q$ means a polarization direction that is parallel to the R.A. axis (IAU convention; \citealt{Contopoulos1974tiau.book.....C}). The level of negative Stokes~$Q$ (in absolute value) compared to the level of positive Stokes~$Q$ within the image appear similar at $3.1$~mm, $2.1$~mm, and $1.5$~mm. At $1.3$~mm, the negative region is stronger (in absolute value) than the positive region. At $870 \mu$m, the alternating positive and negative Stokes~$Q$ differs slightly from \cite{Stephens2017} which showed a largely negative region across most of the disk. For the VLA Q-Band, the point detected to the west is positive and that to the south is negative, which matches the results from the ALMA wavelengths. 

% The difference can be attributed to the difference in angular resolution. Actually, convolving to a large beam doesn't recreate the mostly negative Stokes Q... that's weird

In contrast to Stokes~$I$ and $Q$, Stokes~$U$ (Fig.~\ref{fig:obs_stokes_1}, third column) clearly changes in a rather smooth and consistent manner. 
Starting at $7.1$~mm, the two points detected to the northeast and southwest are both negative, Recall that $+U$ means a polarization direction that is $45^{\circ}$ East-of-North and $-U$ means a polarization direction that is $135^{\circ}$ East-of-North. The negative points match the much better detected $3.1$~mm Stokes~$U$ image, which has negative Stokes~$U$ regions along the northeast and southwest, while the positive Stokes~$U$ regions are along the northwest and southeast. The positive and negative regions are similar in absolute brightness. At $2.1$~mm, the distribution of negative and positive Stokes~$U$ is similar to $3.1$~mm, but the negative region is weaker (in absolute value) than the positive region. A similar trend follows through $1.5$~mm and $1.3$~mm until the negative region becomes absent at $870 \mu$m with positive Stokes~$U$ covering the whole disk. The gradual change of Stokes~$U$ is the main reason why the distribution of $P$ and the polarization directions change smoothly and systematically across wavelengths. 

The polarization fraction, $p$, also changes gradually (Fig.~\ref{fig:obs_stokes_1}, last column). At $3.1$~mm, $p$ is larger away from the center, as expected given the low $P$ at the center in Fig.~\ref{fig:complete_wcs}b. In addition, $p$ is largely azimuthally uniform, varying from $\sim 1.7 \%$ to $2.5\%$, with a slightly larger value along the disk minor axis. At $2.1$~mm, $p$ is also low at the center and the azimuthal variation is also not obvious. From $1.5$~mm to $0.87$~mm, there are two $p$ peaks along the major axis, while $p$ appears consistently lower along the minor axis. The median $p$ from Q-Band to Band 7 (Table~\ref{tab:basic_image_parameters}) drops monotonically from $\sim 7\%$ to $\sim 0.9\%$.

The smooth transition of the morphology of the polarization direction and $p$ can be explained by optical depth effects of scattering, aligned grains \citep{Lin2022}. At the longer wavelength where the disk is optically thinner, the polarization is mainly dominated by polarization from toroidally aligned prolate grains to produce the azimuthally oriented pattern. At shorter wavelengths with larger optical depth, the polarization becomes dominated by scattering which gives a uniform polarization direction parallel to the disk minor axis. The morphologies of the new Bands~4 and 5 polarization images fit surprisingly well with the trend established from the longer (Band~3) and shorter (Bands~6 and 7) wavelength data, indicating that their differences are caused by a relatively simple piece of physics, which we identify as the optical depth effect.

Unlike the smooth morphological transitions in the linear polarization, Stokes~$V$ varies with wavelength more erratically. No discernible Stokes~$V$ emission was detected in the ALMA data that exceeded the anticipated levels attributable to instrumental effects. Since Stokes~$V$ is not the focus of this paper, we leave the results in Appendix~\ref{sec:stokes_V}.

\section{Polarization Analysis} \label{sec:analysis}
From Section.~\ref{sec:results}, we find a systematic transition of the polarization angle from being uniformly parallel to the disk minor axis at the shortest wavelength to being azimuthally oriented around the center at the longest wavelength. To quantify the transition, we follow the technique developed from \cite{Lin2022} which disentangles the azimuthal variation of polarization from a constant component. The technique relies on the approximation that scattering mainly produces a constant polarization due to inclination, thermal polarization produces the azimuthal variation, and both quantities add linearly based on polarized radiation transfer calculations in a simplified plane-parallel geometry.

In the following, Section~\ref{sec:principal_frame_view} describes a particular reference frame to analyze the Stokes~$Q$ and $U$ in a standardized way. Using Stokes~$Q$ and $U$ instead of $P$ is beneficial since they retain the information on both the level of polarization and the direction. Section~\ref{sec:linear_decomposition} introduces the linear decomposition method and measures the spectrum of the scattering component and thermal component. We also find an intriguing asymmetry along the disk minor axis, which we analyze in Section~\ref{sec:near_far_side_asymmetry}. 

\subsection{Principal Frame View} \label{sec:principal_frame_view}

\begin{figure}
    \centering
    \includegraphics[width=0.8\columnwidth]{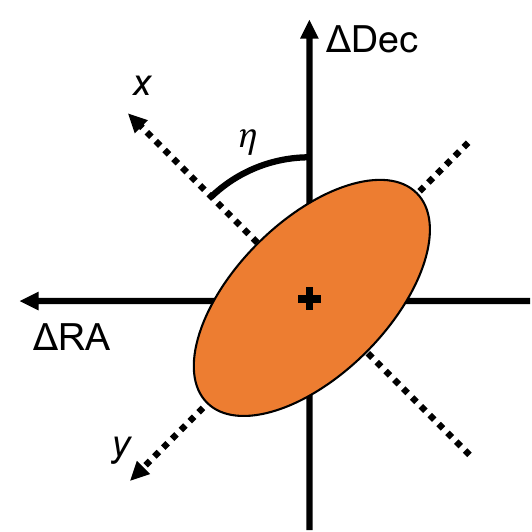}
    \caption{Schematic of the defined orientation of the disk and the principal frame with respect to the plane of sky. $\Delta$RA and $\Delta$Dec (solid arrows) are the coordinates in RA and Dec with respect to the disk center (central cross). The principal frame is defined from the $x$ and $y$ coordinates (dashed arrows). $\eta$ is the angle of the $x$-axis from the $\Delta$Dec axis (East-of-North). }
    \label{fig:sky_schematic}
\end{figure}

Stokes~$Q$ and $U$ depend on the orientation of the image frame. We define an image frame with coordinates $x$ and $y$, such that the $x$- and $y$-axes are along the disk minor and major axes, respectively. Since there is a 180$^{\circ}$ ambiguity in the direction of $x$ (and, likewise, $y$), we arbitrarily fix the positive $x$-direction to the far side of the disk. The positive $y$-direction is 90$^{\circ}$ (East-of-North) from that. Fig.~\ref{fig:sky_schematic} is a schematic that shows the $x$ and $y$ coordinates with respect to the disk minor and major axes. We use the term ``principal'' frame, since it is oriented along the principal axes (i.e., major and minor axes) of an inclined axisymmetric disk. 

The Stokes~$Q'$ and $U'$ defined in the principal frame (denoted with a prime) follow the usual definition from the Institute of Electrical and Electronics Engineering (IEEE Standard 211, 1969) which is the basis of the IAU convention \citep{Contopoulos1974tiau.book.....C, Hamaker1996A&AS..117..137H, Hamaker1996A&AS..117..161H}. 
Let $\phi$ be the angle in the image plane from the positive $x$-axis that increases in the counter-clockwise direction (in the same direction as going East-of-North). Positive $Q'$ is polarization along $x$ ($\phi=0^{\circ}$) and positive $U'$ is polarization along the bisectrix of the positive $x$ and $y$ axes ($\phi=45^{\circ}$). Note that the coordinate system is different from the definition adopted in \cite{Lin2022}, and we provide the derivation of the principal frame that strictly follows the IEEE definition. This frame is motivated by the fact that the scattering of an inclined disk largely produces unidirectional polarization parallel to the disk minor axis, which would show as positive Stokes~$Q'$ and zero Stokes~$U'$.

Under this definition, Stokes~$Q'$ and $U'$ are related to the Stokes $Q$ and $U$ in the original sky frame with a simple rotation. We use $\Delta$RA and $\Delta$Dec as the coordinates in the original sky frame with respect to the center of the disk. Fig.~\ref{fig:sky_schematic} also shows the relation between the sky frame to the principal frame. In the sky frame, let $\eta$ be the position angle (East-of-North) of the minor axis of the disk that corresponds to the far side (i.e., the positive direction of the $x$-axis). The coordinates in the principal frame are related to the sky frame by 
\begin{align}
    \begin{pmatrix}
        x \\
        y
    \end{pmatrix}
    = 
    \begin{pmatrix}
        \cos \eta & \sin \eta \\
        - \sin \eta & \cos \eta 
    \end{pmatrix}
    \begin{pmatrix}
        \Delta\text{Dec} \\
        \Delta\text{RA}
    \end{pmatrix}. 
\end{align}
The Stokes~$Q$ and $U$ in the sky frame are related to the Stokes $Q'$ and $U'$ of the principal frame by 
\begin{align}
    \begin{pmatrix}
        Q' \\
        U'
    \end{pmatrix}
    = 
    \begin{pmatrix}
        \cos 2 \eta & \sin 2 \eta \\
        - \sin 2 \eta & \cos 2 \eta 
    \end{pmatrix}
    \begin{pmatrix}
        Q \\
        U
    \end{pmatrix}. 
\end{align}

%The choice of making the horizontal axis along the major axis is not particularly unique for an axisymmetric disk. Choosing the horizontal axis to be along the minor axis or reversing the near/far sides will have the exact same benefits for the rest of the analysis. The chosen orientation simply makes interpretations of a disk as intuitive and simple as viewing a plate placed on the dinner table in daily life. 

The definition of $\eta$ is different from the position angle of the disk major axis that is usually reported. 
The position angle of the disk major axis is $138.02^{\circ}$ based on high angular resolution images from \cite{ALMApartnership2015}. The far side of the disk is to the northeast since the outflow direction is blueshifted to the northeast and redshifted to the southwest \citep{ALMApartnership2015, Yen2017A&A...608A.134Y}. Thus, we have $\eta = 48.02^{\circ}$. 
%The value from \cite{ALMApartnership2015} is largely similar to the position angle of the best-fit 2D Gaussian from Table~\ref{tab:gaussian_fitting}. 

\begin{figure}
    \centering
    \includegraphics[width=0.97\columnwidth]{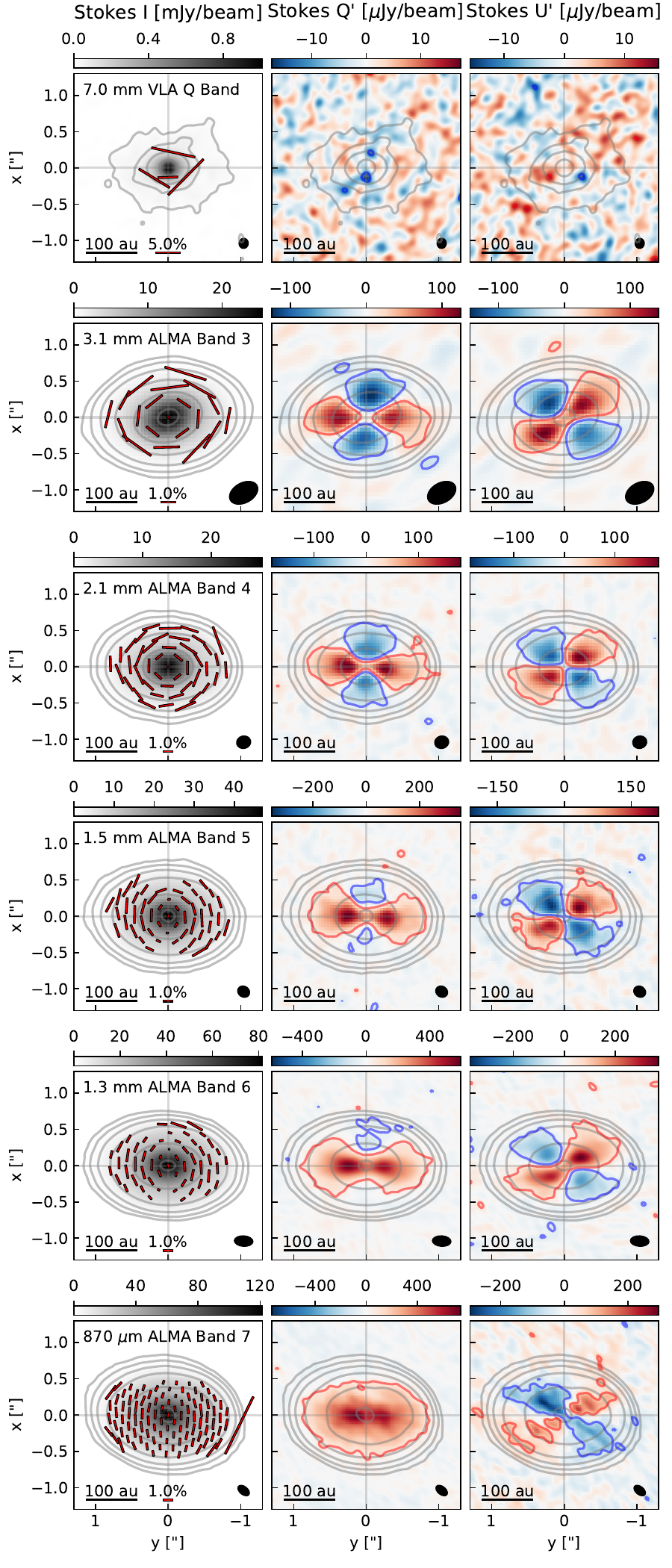}
    \caption{
        Stokes~$I$, $Q'$, and $U'$ at each band where $Q'$ and $U'$ are Stokes $Q$ and $U$ rotated to the principal frame. $I$, $Q'$, and $U'$ images go left to right, while bands Q, 3, 4, 5, 6, and 7 go from the top to the bottom row. The vertical axis ($x$-axis) of the image is along the disk minor axis with $x>0$ defined to be along the far side. The horizontal axis ($y$-axis) is along the disk major axis. The line segments on top of the Stokes~$I$ images represent the polarization direction and the segment length is proportional to $p$ where the scale bar is shown at the bottom. The color scales of Stokes~$Q'$ and $U'$ are plotted such that the white corresponds to the zero level. The $-3\sigma$ and $3\sigma$ levels are marked by blue and red contours respectively. The synthesized beam is represented as a black ellipse to the lower right of each plot.
    }
    \label{fig:obs_stokes_principal}
\end{figure}

Stokes~$Q'$ and $U'$ images are shown in Fig.~\ref{fig:obs_stokes_principal}. For direct comparison, we also show Stokes~$I$ in the principal frame, which is equal to Stokes~$I$ in value but simply rotated. We can easily understand the multiwavelength transition in this frame (at least for the well-detected ALMA images). Across wavelength, from Band~3 ($3.1$~mm) to Band 7 ($870 \mu$m), Stokes~$Q'$ shifts from a petal pattern with alternating signs in each quadrant to an image that is entirely positive. Stokes~$U'$ is mostly zero along the principal axes and the petal pattern with alternating signs does not change with wavelength as Stokes~$Q'$ does. 
Note that Stokes~$Q$ and $U$ images (Fig.~\ref{fig:obs_stokes_1}) appear ``swapped'' with Stokes~$Q'$ and $U'$ images (Fig.~\ref{fig:obs_stokes_principal}) only because $\eta$ for HL Tau happens to be near $45^{\circ}$ and is not generally true for different disks. 

\subsection{Linear Decomposition} \label{sec:linear_decomposition}

\subsubsection{Methodology}
Solving the polarized radiation transfer equation including polarized thermal emission and scattering of elongated grains self-consistently is notoriously challenging \citep[e.g.][]{Steinacker2013ARA&A..51...63S}. Nevertheless, as demonstrated by \cite{Lin2022}, the problem simplifies significantly in a plane-parallel slab. Since the dust layer responsible for the HL Tau (sub)millimeter continuum is geometrically thin \citep{Pinte2016ApJ...816...25P}, one can approximate each local patch of the dust disk as a plane-parallel slab.

In addition, \cite{Lin2022} found that, when the optical depth is less than of order unity, the polarization fraction is approximately a linear addition of polarization due to thermal emission of the elongated grain without scattering and polarization due to scattering of a volume-equivalent sphere when the shape of the grain is nearly spherical. When the optical depth is large, the resulting polarization fraction is largely determined by scattering alone. The approximation enables us to sidestep complications arising from the full disk geometry and (uncertain) grain opacities and directly estimate the contributions from scattering and thermal emission from the azimuthal variation. We limit the model to the ALMA Bands since the polarization is better detected around the full azimuth. 
%Moreover, from our modeling of the ALMA bands, we make predictions of the polarized emission at 7 mm which are surprisingly consistent with the marginal detection at Q-Band.

For clarity, we provide the essential derivation with the appropriate convention adopted in this work (see \citealt{Lin2022} for the original derivation). Assuming a prolate grain in the dipole limit, the polarization purely from thermal emission is \citep{Lee1985ApJ...290..211L, Yang2016MNRAS.460.4109Y}:
\begin{align} \label{eq:dipole_approximate_p_grain_frame}
    p(\theta_{g}) = \frac{ p_{0} \sin^{2} \theta_{g} }{ 1 - p_{0} \cos^{2} \theta_{g} } \approx p_{0} \sin^{2} \theta_{g}
\end{align}
where $\theta_{g}$ is the viewing angle from the axis of symmetry of the grain ($\theta_{g}=0^{\circ}$ means the grain is seen pole-on). Recall that the use of a lowercase refers to a quantity that is related to the polarization fraction (normalized by $I$). We define $p_{0}$ as the intrinsic polarization which is the polarization of the grain seen edge-on ($\theta_{g}=90^{\circ}$) and is the maximum polarization possible just from the shape. The approximation to the right-hand-side of Eq.~(\ref{eq:dipole_approximate_p_grain_frame}) applies because $p_{0} \ll 1$.

Dichroic extinction attenuates the polarization as optical depth increases \citep[e.g.][]{Hildebrand2000PASP..112.1215H, Lin2022}. Since $p_{0} \ll 1$, the resulting polarization remains $\propto p_{0} \sin^{2} \theta_{g}$, so we express the thermal polarization as
\begin{align} \label{eq:thermal_polarization_approximation}
    t_{p}(\theta_{g}) = t \sin^{2} \theta_{g}
\end{align}
where $t$ is $p_{0}$ attenuated by optical depth. 
The explicit dependence of $t$ on optical depth can be complicated and is beyond the scope of this paper, but the usefulness of Eq.~(\ref{eq:thermal_polarization_approximation}) is in separating the optical depth attenuation part from the part that only depends on the viewing angle (which gives the azimuthal variation as we see below). 

\begin{figure}
    \centering
    \includegraphics[width=\columnwidth]{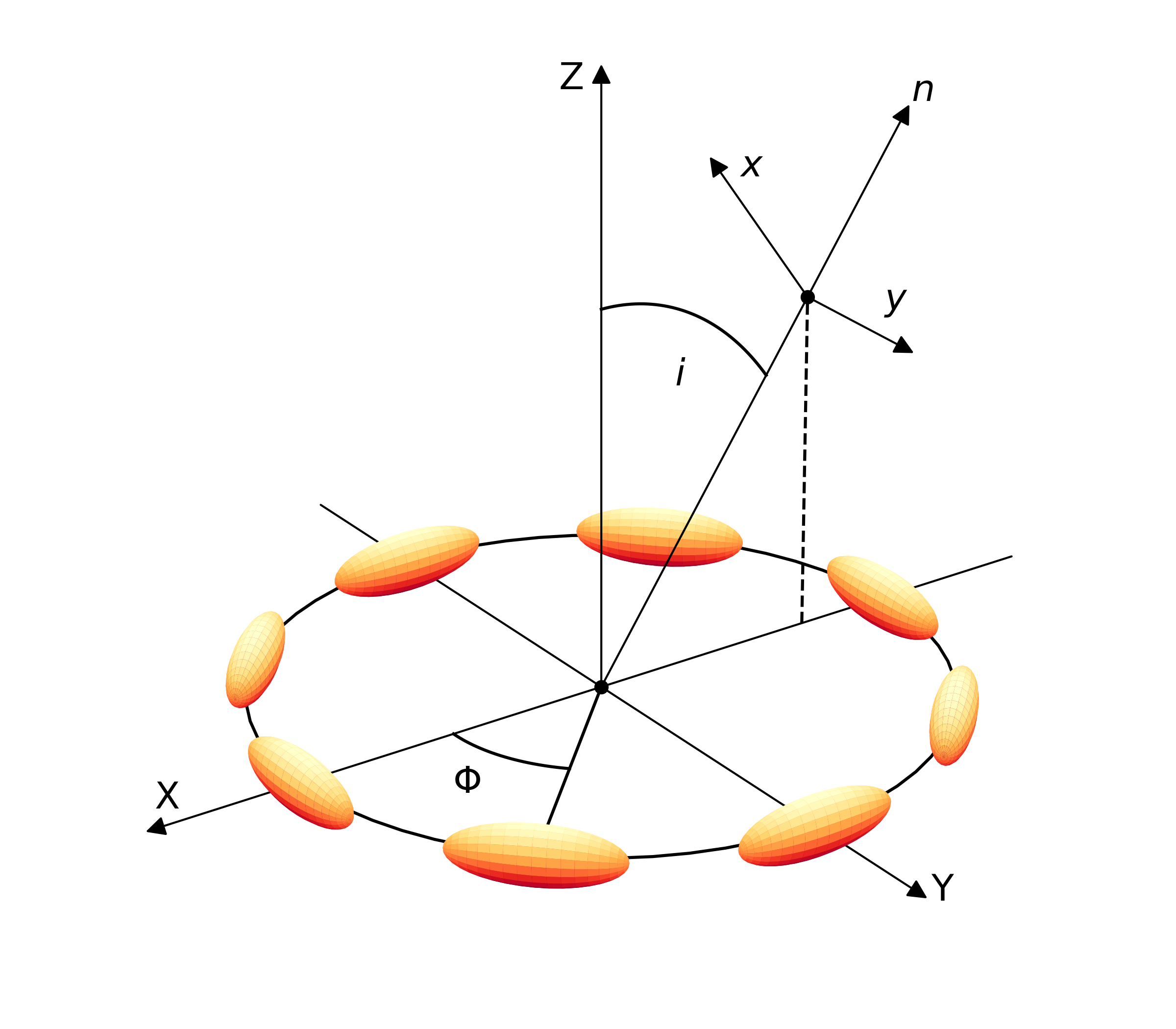}
    \caption{
        Schematic of a disk with toroidally aligned prolate grains in relation to the observer. The $X$ and $Y$-axes form the disk midplane and $Z$ is the rotation axis of the disk. $i$ is the inclination to the observer. $n$ is the direction to the observer, and the $x$- and $y$-axes form the principal frame. $\Phi$ is the azimuthal angle in the disk midplane. The orange prolates represent the aligned grains.
    }
    \label{fig:system_to_sky}
\end{figure}

Next, we consider an inclined axisymmetric disk demonstrated in Fig.~\ref{fig:system_to_sky}. Let $Z$ be the rotation axis of the disk and $\hat{n}$ be a unit vector directed to the observer. The inclination $i$ is the angle between $Z$ and $\hat{n}$. $X$ and $Y$ are axes in the disk midplane such that $X$ is coplanar to $Z$ and $\vect{n}$. We define $\Phi$ as the azimuthal angle in the disk midplane from the $X$-axis without loss of generality since the disk is assumed to be axisymmetric. The alignment axes of the prolate grains are in the disk midplane and in the azimuthal direction. Based on the definition of the principal frame in Sec.~\ref{sec:principal_frame_view}, $x$ is in the $XZ$-plane. For convenience, we define $\phi$ as the azimuthal angle in the image plane from the $x$-axis.

Let $q'\equiv Q'/I$ and $u'\equiv U'/I$ (i.e., normalized $Q'$ and $U'$ in the principal frame). Depending on the location along the azimuth, the viewing angle $\theta_{g}$ varies and gives the azimuthal variation seen in the image. Contribution to $q'$ and $u'$ from thermal emission is given in Appendix~\ref{sec:thermal_polarization_in_principal_frame}. The polarization from the scattering component, which we denote as $s$, is largely constant of azimuth and only contributes to $q'$ since the inclination-induced polarization is always parallel to the disk minor axis. Adding the thermal component and scattering component together, we get
\begin{align} 
    q' &= s + t (\cos^{2} i \sin^{2} \Phi - \cos^{2} \Phi) \label{eq:azimuthal_q_linear_decomposition} \\
    u' &= - t \cos i \sin 2\Phi \label{eq:azimuthal_u_linear_decomposition}. 
\end{align}

%We sample the azimuthal profile in constant steps of sizes equal to the beam size, because the each finite resolution element is linear in the image plane. In this case, $\phi$ is constant... wait that's not true, but $\Phi$ become non-linear and is related to $\phi$ by through simple geometrical arguments: 
%\begin{align} 
%    \Phi = \text{arctan2}(\sin \phi \cos i, \cos \phi). 
%\end{align}

Using Eq.~(\ref{eq:azimuthal_q_linear_decomposition}) and (\ref{eq:azimuthal_u_linear_decomposition}) we fit the azimuthal profile of $q'$ and $u'$, respectively, at 100~au first for each band. The chosen radius is $\sim 2$ beams away from the center for the Band~3 image, which has the poorest resolution, to minimize the effects of beam convolution, but is also within a range with enough signal-to-noise for all five bands. We conduct the same process for other radii below. Sampling the azimuthal profile uses steps equal to the geometric average of the beam size.

We use \textit{emcee}, a Monte Carlo Markov Chain sampling code \citep{ForemanMackey2013PASP..125..306F}, to find the best-fit values and uncertainties of $s$ and $t$ at each wavelength. We use 32 walkers and a total of 2500 steps. We ignore the first 500 steps to obtain the posterior probability distribution. Modifying the walking parameters does not significantly change the results. The best-fit values are determined from the median of the marginalized distribution, and the $1\sigma$ uncertainties use the 16th and 84th percentile. We show the two-dimensional posterior probability distribution derived from \textit{emcee} in Appendix~\ref{sec:corner_plots}.

\begin{figure*}
    \centering
    \includegraphics[width=\textwidth]{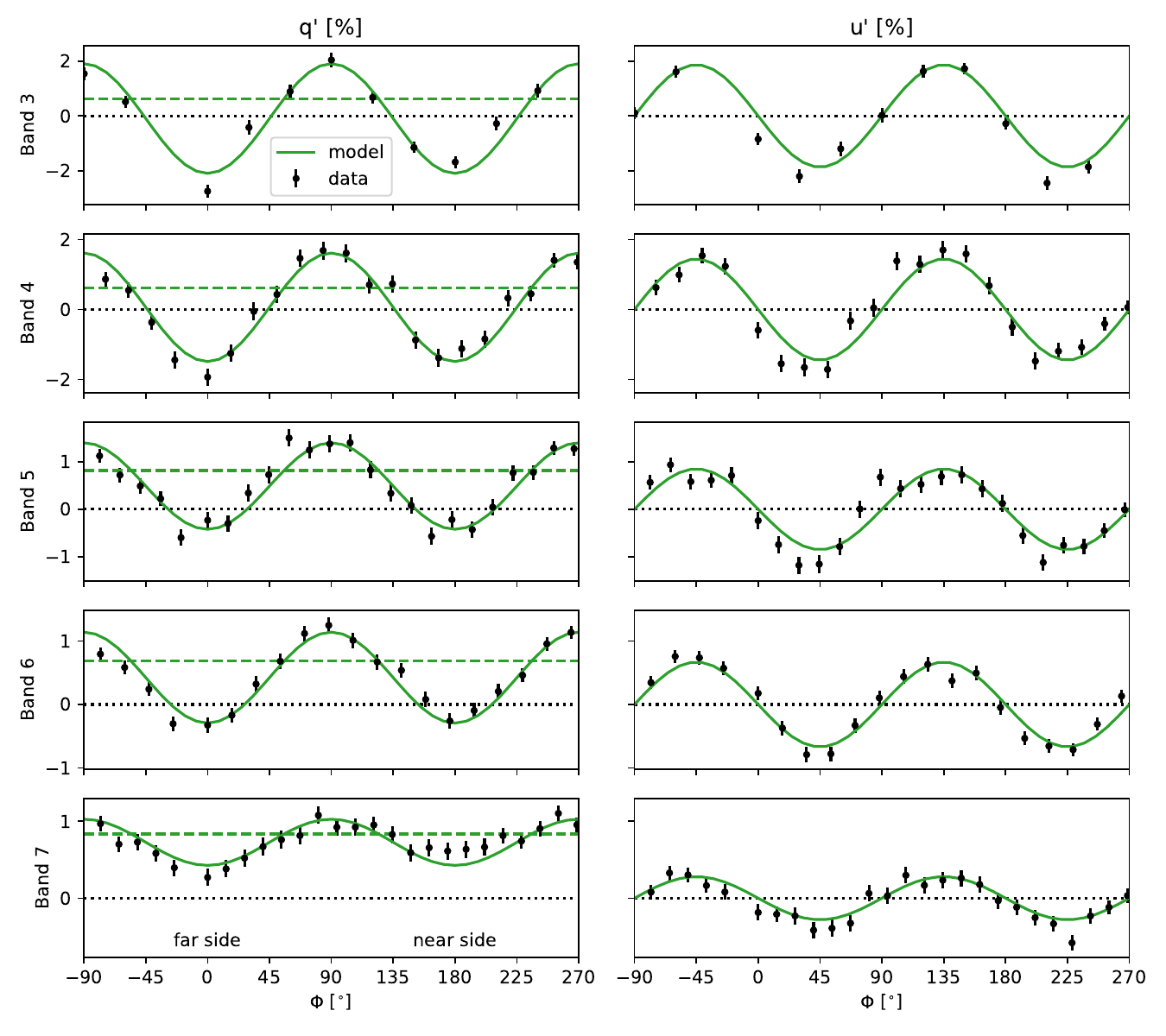}
    \caption{
        The azimuthal variation of $q'$ and $u'$ (left and right columns) for each ALMA Band (from top to bottom). The data are shown in black dots with error bars corresponding to the statistical uncertainty. The green curves are the best-fit model curves, and the green horizontal dashed curve in the left panels is the best-fit $s$ component. The black horizontal dotted curve is the zero line. $\Phi$ is plotted from -90$^{\circ}$ to 270$^{\circ}$ to better see the complete near and far sides. 
    }
    \label{fig:fit_azimuthal}
\end{figure*}

\subsubsection{Results}

Fig.~\ref{fig:fit_azimuthal} shows the best-fit curve of the model compared to the sampled observational data points for the high signal-to-noise ALMA observations. We find that the linear decomposition model describes all five bands, in both $q'$ and $u'$, remarkably well considering the simplicity of the model. While this was already shown for the same Bands 3 and 6 data with just a difference in the self-calibration and imaging procedure \citep{Lin2022}, it is reassuring to see that the new Bands 4, 5, and 7 data follow the same pattern, which adds weight to the validity of the simple decomposition technique. Intriguingly, $q'$ of the near side ($\Phi \in [90^{\circ}, 270^{\circ}]$) appears slightly, but systematically larger than the best-fit model, while $q'$ of the far side  ($\Phi \in [-90^{\circ}, 90^{\circ}]$) appears systematically lower, indicating another, more secondary effect is also at play. We discuss the near-far side asymmetry in Section~\ref{sec:near_far_side_asymmetry}.

\begin{figure*}
    \centering
    \includegraphics[width=0.85\textwidth]{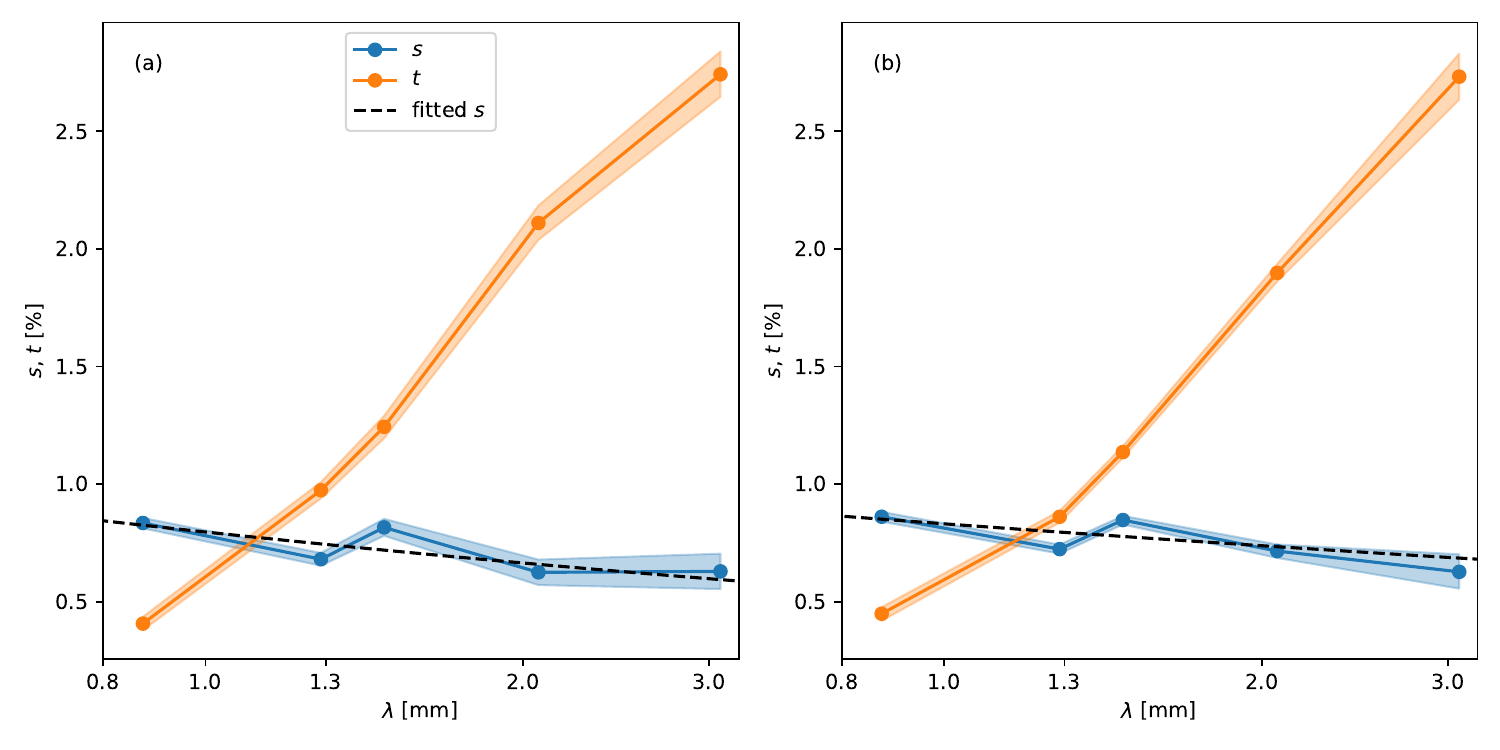}
    \caption{
        The spectrum of $s$ and $t$: $s$ is the level of polarization from scattering, while $t$ is the intrinsic polarization from aligned grains attenuated by optical depth. Panel (a): The results from fitting the data at their native resolution. Panel (b): The results from fitting the data after convolving Bands 4 to 7 with smaller beam sizes to the Band~3 beam size. The blue and orange curves are $s$ and $t$, while the shaded regions represent the $1\sigma$ uncertainty from the fit. The dashed line is the best-fit power-law curve to the $s$ spectrum.
        }
    \label{fig:st_spectrum}
\end{figure*}

Fig.~\ref{fig:st_spectrum}a shows the best-fit $s$ and $t$ as a function of wavelength including the uncertainties estimated from \textit{emcee}. Evidently, the contribution from thermal polarization, $t$, monotonically increases with increasing wavelength. The behavior is what we expect from a decrease in optical depth as the dust opacity decreases towards longer wavelengths \citep{Hildebrand2000PASP..112.1215H, Yang2017MNRAS.472..373Y}.

The contribution from scattering, $s$, slowly decreases with increasing wavelength in general with the exception of $s$ at $\lambda$=1.5~mm (Band~5) which appears slightly larger than $s$ at $\lambda=1.3$~mm (Band~6). 
To describe the spectrum of the scattering component, we fit a power-law in the form of $a (\lambda / 1 \text{mm})^{b}$. We again use \textit{emcee} and obtain $a \sim 0.796 \pm 0.016$~$\%$ and $b \sim -0.26 \pm 0.06$. The two-dimensional posterior distribution is also included in Appendix~\ref{sec:corner_plots}. The overall decrease of $s$ (negative $b$) is what we expect due to decreasing optical depth. How slowly $s$ decreases may depend on the optical depth, opacity index, grain size, and porosity which we discuss in Section~\ref{sec:discussion}.

The slight increase of $s$ at Band~5 could be due to the maximized scattering (inclination-induced) polarization when the optical depth is of order unity \citep{Yang2017MNRAS.472..373Y, Lin2022}. Multiwavelength continuum ray-tracing from \cite{Pinte2016ApJ...816...25P} showed that, at a radius of 100~au, the optical depths at Band~3 (2.9~mm) and 6 (1.3~mm) are $\sim 0.4$ and $0.3$, respectively, though the modeling did not consider scattering. Nevertheless, including scattering, \cite{CarrascoGonzalez2019ApJ...883...71C} obtained optical depths of $\sim 1$ and $\sim 3$ at Bands~4 (2.1~mm) and 6 (1.3~mm), respectively, at the same radius. Band~5 (1.5~mm), being in between the wavelengths considered in the previous two studies, appears likely to have an optical depth necessary to maximize the inclination-induced polarization.

We note that when comparing properties across wavelengths, it is preferable to use the same spatial resolution. Thus, we conduct the same procedure at the same radius, but with all the data convolved to the same resulting beam size using the CASA \textit{imsmooth} task. We use the beam size from Band~3 which is the largest among the five bands.

The resulting $s$-spectrum (Fig.~\ref{fig:st_spectrum}b) is comparable to the original profile, which is reasonable since scattering polarization is largely unidirectional and the averaging effects from a moderately larger beam will not introduce significant cancellations. Indeed, by fitting the $s$ spectrum, we get $a=0.831 \pm 0.016 \%$, $b=-0.17 \pm 0.05$, which is comparable to the values obtained in the previous case. 
The resulting $t$-spectrum (Fig.~\ref{fig:st_spectrum}a) remains monotonically increasing with wavelength and does not change significantly from Fig.~\ref{fig:st_spectrum}a) However, the slight drop in $t$ (most clearly seen at Band 4) after convolution is because of the beam cancellation of its azimuthal polarization.

We conduct the same process at each radius to obtain the radial dependence of $s$ and $t$ at each wavelength. To maximize the benefits of the high angular resolution, we fit the azimuthal profile at the images with their original resolution. The minimum radius is chosen such that we have at least eight points to fit the sinusoidal curve. The maximum radius cuts off where there is not enough $3\sigma$ detection around the azimuth. 
% we don't need Nyquist sampling, because that's more like to determine the frequency. We know the frequency, so we just need the amplitude... is that more stringent or less? 

\begin{figure}
    \centering
    \includegraphics[width=\columnwidth]{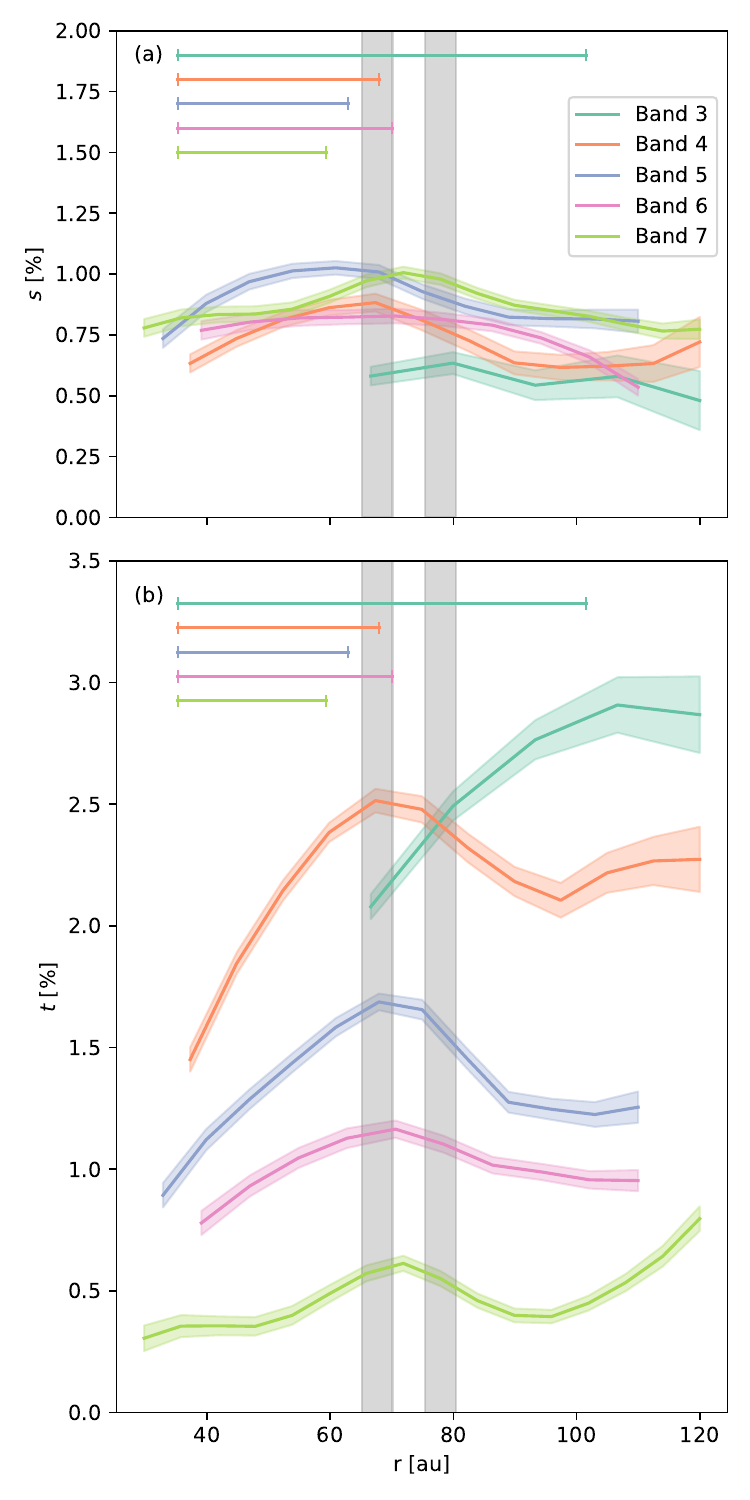}
    \caption{
        The radial profiles of $s$ (panel a) and $t$ (panel b) in units of percent compared across bands. The shaded region corresponds to the $1\sigma$ uncertainty from the fit. The horizontal lines correspond to the beam FWHM projected along the disk minor axis and the colors match the legend. The shaded vertical bars mark the particularly deep gaps at 68 and 78~au. 
    }
    \label{fig:st_radius}
\end{figure}

Fig.~\ref{fig:st_radius} shows the resulting $s$ and $t$ as a function of radius. For each wavelength, $s$ appears largely constant with radius though there is a hint of stronger $s$ at inner radii. However, there is a large scatter and there are no obvious coherent radial changes across wavelengths. On the other hand, the radial profiles of $t$ appear to share a few features across wavelengths. The drop towards the inner radius ($< 50$~au) is likely due to beam averaging which artificially decreases the azimuthal variation from thermal polarization by aligned grains. For Bands 4 to 7, $t$ appears to peak at $r\sim70$~au and drop to a minimum at $r\sim 90$~au. $t$ of Band~3 does not share a similar variation due to the much larger beam size. The consistent variation between Bands 4 to 7 is likely due to the underlying substructure. The peak at $r\sim 70$~au appears to coincide with the two close and deepest gaps at $68$ and $78$~au \citep{ALMApartnership2015}. The minimum at $r\sim 90$~au coincides with the ring at $86$~au. $t$ reaches a peak in the low surface density region and becomes a minimum in the high surface density, which is what we expect from optical depth effects of polarization from aligned grains \citep{Hildebrand2000PASP..112.1215H}. 

\subsection{Near-far Side Asymmetry} \label{sec:near_far_side_asymmetry}

From Section~\ref{sec:linear_decomposition}, there appears to be an asymmetry between the near side and far sides of the disk which we explore in this section. The asymmetry is visually evident from the Stokes~$Q'$ image in Fig.~\ref{fig:obs_stokes_principal} (second column) and also from $P$ in Fig.~\ref{fig:obs_stokes_1}.

To make a direct comparison, we use a cut along the disk minor axis. The cut uses a slit along the minor axis with a finite width equal to the beam size and averages the Stokes parameters in the principal frame along the width. To see the difference between the Stokes~$I$ of the near and far sides, we define the fractional difference, $f$, as the ratio between the difference of Stokes~$I$ in the near and far sides to their average 
\begin{align} \label{eq:stokes_I_near_far_asymmetry}
    f \equiv \frac{ I_{\text{near}} - I_{\text{far}} }{ (I_{\text{near}} + I_{\text{far}}) / 2 }
\end{align}
where $I_{\text{near}}$ and $I_{\text{far}}$ are the Stokes~$I$ along the near side and far side, respectively. The uncertainty of $f$ is estimated through error propagation. In the principal frame, $P$ is well represented by Stokes~$Q'$, since Stokes~$U'$ is $\sim 0$ and using Stokes~$Q'$ retains the sign to represent the direction of polarization. Likewise, we show $q'$ which fully represents the polarization fraction while retaining the direction of polarization. 

\begin{figure*} 
    \centering
    \includegraphics[width=\textwidth]{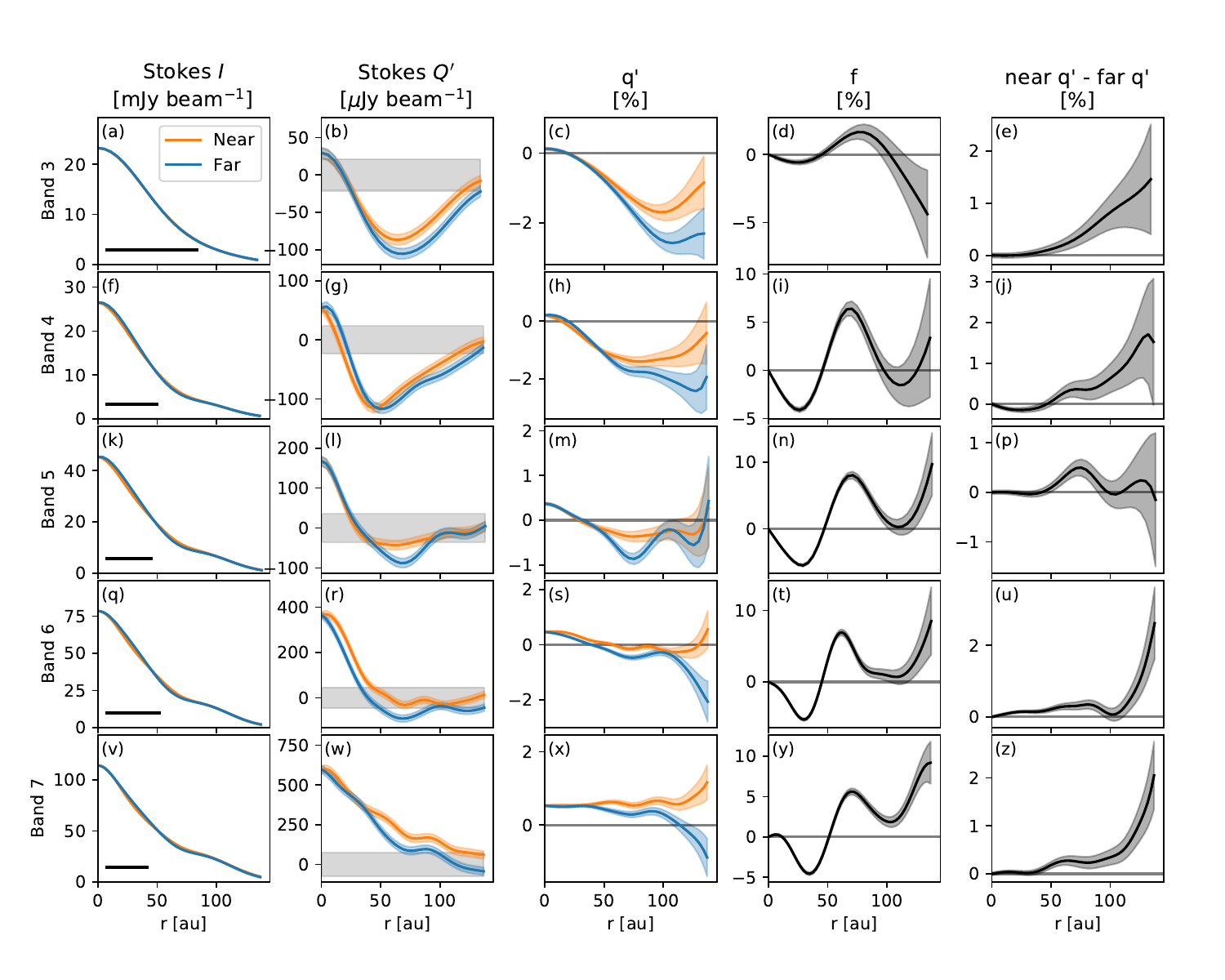}
    \caption{
        The profiles along the disk minor axis from the left column to the right column are Stokes~$I$, $Q'$, $q'$, f, and the difference in $q'$. The top to bottom rows are from Bands 3 to 7. The shaded region represents the uncertainty. The horizontal axis is the deprojected radius in au. The horizontal black line segment to the lower left in the first column is the deprojected beam FWHM. 
    }
    \label{fig:nearfar_sides}
\end{figure*}

Fig.~\ref{fig:nearfar_sides} shows that the Stokes~$I$ appears rather symmetric across the disk minor axis from Bands 3 to 7 (Fig.~\ref{fig:nearfar_sides}, first column). The fractional differences are $\sim 5\%$ across bands based on $f$ (Fig.~\ref{fig:nearfar_sides}, fourth column). In contrast, Stokes~$Q'$ is visibly asymmetric and, in most cases, the Stokes~$Q'$ of the near side is greater than the Stokes~$Q'$ of the far side. Specifically, we can see this case at $r>25$~au for Band~3 (Fig.~\ref{fig:nearfar_sides}b), at $r>50$~au for Band~4 (Fig.~\ref{fig:nearfar_sides}g), at $50<r<80$~au for Band~5 (Fig.~\ref{fig:nearfar_sides}l), at $r<80$~au for Band~6 (Fig.~\ref{fig:nearfar_sides}r), and at $r>40$~au for Band~7 (Fig.~\ref{fig:nearfar_sides}w). In fact, the only region where the Stokes~$Q'$ of the near side is less than that of the far side is at $r<50$ for Band~4 (Fig.~\ref{fig:nearfar_sides}f). 

The symmetric Stokes~$I$ and asymmetric Stokes~$Q'$ results in $q'$ with similar regions of asymmetry as Stokes~$Q'$. The rightmost column of Fig.~\ref{fig:nearfar_sides} shows the difference between the near side $q'$ and the far side $q'$. Note that this is not the \textit{fractional} difference, like that used for Stokes~$I$. At small radii, the difference is small and largely consistent with no difference. At regions with more confident detection, $r \sim 70$~au, the near side $q'$ is larger than the far side $q'$ by $\sim 0.3\%$ across bands. At even larger radii, the difference increases to $\sim 1\%$, but with less certainty. 
%\red{fractional change is large for q}

No asymmetry along the minor axis was found in \cite{Stephens2017} for Band~7. However, the presented Band~7 with better angular resolution may have made it easier to detect. In addition, the consistent offset across wavelengths strengthens the case that the asymmetry is real. We discuss the origin of the near-far side asymmetry in Section~\ref{sec:origin_of_nearfar_side_asymmetry}. 

\subsection{Q-Band consistency with toroidally aligned grains}

\begin{table*}
    \centering
    \begin{tabular}{c c c c c c c c c c c c}
        \hline
        $\Delta$RA & $\Delta$Dec & $I$ & $P$ & $P$ SNR & $p$ & $\sigma_{\text{pf}}$ & $\chi$ & $\sigma_{\chi}$ & $\Delta \chi$ & $t$ & $\sigma_{t}$\\
        $\arcsec$ & $\arcsec$ & mJy/beam & $\mu$Jy/beam & & $\%$ & $\%$ & $^{\circ}$ & $^{\circ}$ & $\sigma_{\chi}$ & $\%$ & $\%$ \\
        (1) & (2) & (3) & (4) & (5) & (6) & (7) & (8) & (9) & (10) & (11) & (12) \\
        \hline
        -0.093 & -0.090 & 0.417 & 17 & 4.3 & 4.0 & 0.9 & 140 & 7 & 0.46 & 4.1 & 0.9 \\
        0.112 & 0.195 & 0.154 & 14 & 3.5 & 9 & 3 & 126 & 8 & -0.43 & 9 & 3 \\
        -0.256 & 0.096 & 0.128 & 13 & 3.3 & 10 & 3 & 3 & 9 & 0.36 & 15 & 5 \\
        0.022 & -0.238 & 0.182 & 13 & 3.4 & 7 & 2 & 107 & 9 & 0.08 & 10	& 3 \\
        \hline
    \end{tabular}
    \caption{
        Properties of the detected polarization values at Q-Band. Columns 1 and 2: The RA and Dec relative to the adopted center of the disk, respectively. Columns 3 and 4: The Stokes~$I$ and $P$. Column 5: The signal-to-noise ratio of $P$. Columns 6 and 7: The polarization fraction and the uncertainty. Columns 8 and 9: The polarization angle and its uncertainty. Column 10: The observed $\chi$ with respect to the model $\chi$ in units of $\sigma_{\chi}$. Columns 11 and 12: The expected intrinsic polarization $t$ of the grain after deprojection of $p$ and its uncertainty, respectively.  
    }
    \label{tab:Qband_vectors}
\end{table*}

Using the longest wavelength, VLA Q-band data, we check if the polarization angles, $\chi$, are consistent with toroidally aligned prolate grains. Table~\ref{tab:Qband_vectors} lists the measurements of each detected polarization vector, like the spatial location, $\chi$, and $\sigma_{\chi}$. Following Sections~\ref{sec:principal_frame_view} and \ref{sec:linear_decomposition}, we can deproject the location of the detected vectors and derive the expected $\chi$ from toroidally aligned prolate grains. We use Eq.~(\ref{eq:azimuthal_q_linear_decomposition}) and (\ref{eq:azimuthal_u_linear_decomposition}), but assume $s=0$ to derive the $\chi$ in the principal frame and rotate it to the sky frame. 

\begin{figure}
    \centering
    \includegraphics[width=\columnwidth]{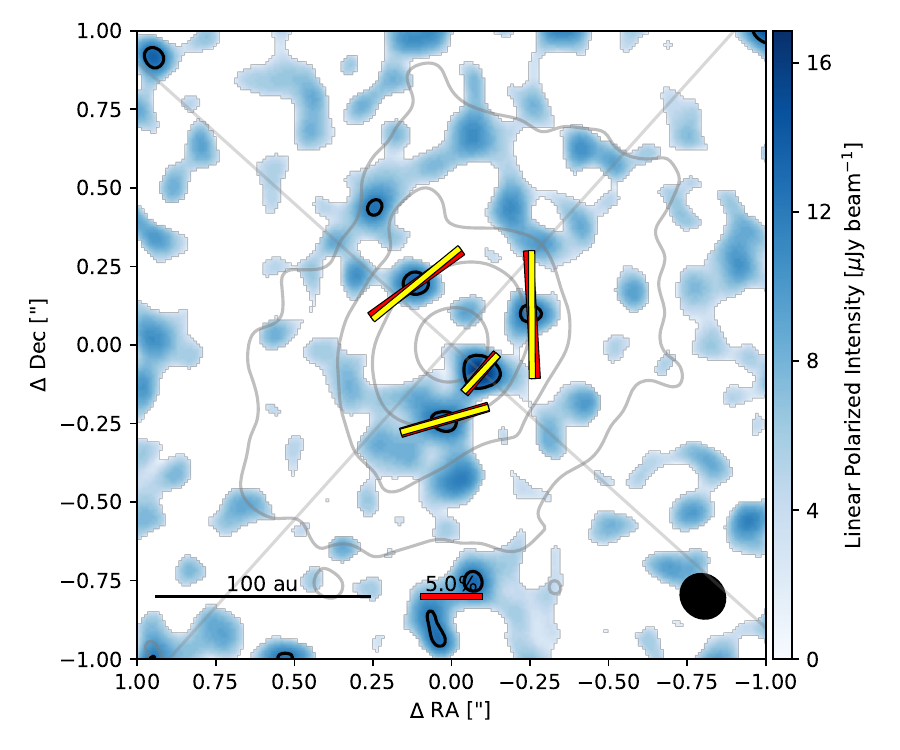}
    \caption{
        Comparison between the observed polarization direction (red vectors) and the expected polarization direction from toroidally aligned prolate grains (yellow vectors). The lengths are made to match the observed $p$ and only the polarization direction should be compared. The color map is $P$ and the grey contours are the Stokes~$I$ in steps of 3, 10, 30, 100, 300, and 1000 $\sigma_{I}$. 
    }
    \label{fig:vla_vs_aligned}
\end{figure}

Fig.~\ref{fig:vla_vs_aligned} shows the Q-Band polarization image compared to the expected polarization direction. The differences with the observed $\chi$ normalized by the $\sigma_{\chi}$ are 0.5, -0.4, 0.4, and 0.08 which means the observed $\chi$ are consistent with toroidally aligned prolate grains. In addition, the probability for random noise to have 4 points within $\pm 1\sigma_{\chi}$ of the expected polarization direction is $\Pi_{i=1}^{4} 2 \sigma_{\chi, i} / 180^{\circ} \sim 7 \times 10^{-5}$, where $\sigma_{\chi,i}$ represents the $\sigma_{\chi}$ of the $i$th detection. Thus, the detections are unlikely due to random noise.

From the deprojected locations, we can estimate the level of $t$ using the observed $p$. We find that the values are $4.1 \pm 0.9 \%$, $9\pm 3\%$, $15 \pm 5 \%$, and $10\pm 3 \%$ (Table~\ref{tab:Qband_vectors}) where the uncertainty is from error propagation with only the uncertainty from $p$. When calculating $p$, we did not remove the free-free component as was done in \cite{CarrascoGonzalez2019ApJ...883...71C} because the free-free emission is only within the central $\sim 40$~mas and the polarization detections are at least $\sim 1$~beam ($0.15\arcsec$) away from the center. Thus, the detected vectors are unlikely contaminated by free-free emission.

From the single vector along the disk minor axis with negative $Q'$ and $q'\sim 4\%$ (Fig.~\ref{fig:obs_stokes_principal}), we find $t\sim 4\%$ since the prolate grain is viewed edge-on ($\theta_{g}=90^{\circ}$). The value is greater than $t$ measured from Band~3, which fits the expectation if Band~3 is optically thicker than at Q-Band. However, that particular vector can be contaminated by scattering, which gives positive $Q'$, or artificially diminished due to beam averaging effects since the vector is located near the center. On the other hand, the two vectors located to the south and to the west of the center (Fig.~\ref{fig:vla_vs_aligned}) are at least one beam away from the center making it less susceptible to beam averaging. Deprojecting the vectors to obtain $t$ gives much higher $\sim 10\%$ and $15\%$. Given that the disk is likely optically thin \citep{CarrascoGonzalez2016ApJ...821L..16C, CarrascoGonzalez2019ApJ...883...71C}, we estimate that the intrinsic polarization of grains, $p_{0}$, should be comparable. The value is unlikely diminished due to scattering because the polarization is detected in Stokes~$U'$ (Fig.~\ref{fig:obs_stokes_principal}, third column) where scattering contributes less. $p_{0}$ of $>10\%$ is much higher than the $2\%$ inferred from the ALMA wavelengths \citep{Lin2022} which could be due to the disk being optically thicker at Band~3 \citep{CarrascoGonzalez2016ApJ...821L..16C, CarrascoGonzalez2019ApJ...883...71C}. We suspect that the intrinsic level of polarization for grains may be a lot higher than that derived from low angular resolution polarization observations at the current ALMA bands. Particularly, the estimated value may be consistent with the inferred $p_{0}$ from the gaps resolved by high angular resolution polarization at Band~7 (Stephens et al., in press). Concrete conclusions for the true $p_{0}$ of grains require higher angular resolution images and/or better data at long wavelengths which is possible from longer VLA integration times, ALMA Band~1 once its polarization capability becomes available, or the ngVLA.

\section{Discussion} \label{sec:discussion}

\subsection{Grain structure}
Our main result is that the new ALMA images at Bands 4, 5, and 7 and the VLA Q-band image are consistent with scattering of grains that are effectively prolate and toroidally aligned. This is in line with previous work using just Bands 3, 6, and 7 \citep{Yang2019MNRAS.483.2371Y, Mori2021ApJ...908..153M, Lin2022}. The evidence comes from the increasing azimuthal variation and the decreasing constant component from the ALMA Bands as the wavelength increases from 0.87~mm to 3.1~mm (Sec.~\ref{sec:linear_decomposition}). Though the few marginally detected polarization vectors at Q-band prohibit an anlaysis of the azimuthal variation, the polarization angles are consistent with the toroidally aligned prolate grains in the optically thin limit as predicted in \cite{Lin2022}. From these results, we discuss the implications of the grain structure.

Past studies of HL Tau using (sub)millimeter multiwavelength Stokes~$I$ images require large, mm-sized grains by constraining the opacity index $\beta \sim 1$ \citep{Kwon2011ApJ...741....3K, Kwon2015ApJ...808..102K}. Even after accounting for scattering and optical depth effects, 
\cite{CarrascoGonzalez2019ApJ...883...71C} used resolved ALMA and VLA observations and inferred $\sim 1$~mm grains. The grain size is in tension with that inferred from polarization studies, which limits the grain size to $\sim$100~$\mu$m \citep[e.g.][]{Yang2016MNRAS.456.2794Y, Kataoka2016ApJ...820...54K}.

In Section~\ref{sec:linear_decomposition}, we measured the $s$-spectrum for the ALMA Bands which traces the effective contribution from scattering as a function of wavelength. At face value, $s$ falls approximately as $\propto \lambda^{-0.2}$. The weak dependence on $\lambda$ across 0.87~mm to 3.1~mm is difficult to explain through spherical grains even after considering optical depth effects \citep{Lin2022}. Dust settling with a spatial distribution of various sizes of spherical grains can reproduce the weak $\lambda$ dependence, the level of polarization, and the multiwavelength Stokes~$I$, but most likely for the inner regions where it is optically thick for HL Tau \citep{Ueda2021ApJ...913..117U}.

Irregularity of grain structure has been shown to alleviate the tension between the grain sizes inferred from scattering-induced polarization and those from the spectral index \citep[e.g.][]{Shen2008ApJ...689..260S, Shen2009ApJ...696.2126S, Tazaki2019ApJ...885...52T, Munoz2021ApJS..256...17M, Lin2023MNRAS.520.1210L}. Indeed, \cite{Zhang2023ApJ...953...96Z} simultaneously modeled the Stokes~$I$ and the weak $\lambda$ dependence of polarization assuming porous grains and found that the grains can be greater than 1~mm depending on the porosity.

Aside from the scattering behavior, the lack of any flip in the underlying thermal polarization direction from aligned grains from 870~$\mu$m to 7~mm also constrains the grain structure. Compact elongated grains produce thermal polarization following the direction along the projected long axis when $\lambda$ is much larger than grain size $a$, or more specifically when $\lambda > 2 \pi a$ (Rayleigh regime). However, when $\lambda$ is comparable to $2 \pi a$ (Mie regime), the (thermal) polarization direction can flip, i.e., change by $90^{\circ}$, and become perpendicular to the projected long axis \citep{Kirchschlager2019MNRAS.488.1211K, Guillet2020A&A...634L..15G}. At face value, the lack of any flip even at our shortest wavelength band implies that the grain size should be smaller than $\sim 140$~$\mu$m. However, such a small grain size is unlikely to explain the weak $\lambda$ dependence of the $s$-spectrum or the level of Stokes~$I$ and $P$ at the VLA wavelengths (see also \citealt{Ohashi2020ApJ...900...81O}). Since grains with sizes of order $\lambda/2\pi$ should contribute the most absorption opacity \citep[e.g.][]{Birnstiel2018ApJ...869L..45B}, the detected thermal polarization at 7~mm would be most efficiently emitted from aligned, 1~mm grains. Including porosity alleviates the strict size constraint of $a < \lambda / 2 \pi$, and opens the possibility of aligned, $\sim 1$~mm grains \citep{Kirchschlager2019MNRAS.488.1211K}.

Based on the weak $\lambda$ dependence of scattering polarization and the lack of any flip in the underlying thermal polarization, we suspect that the grains in HL Tau are porous, and prolate with a maximum size of $\sim 1$~mm. Future efforts to incorporate such grains into detailed modeling for comparison with the multi-wavelength Stokes~$I$ and polarization data will be valuable.

\subsection{Implications on grain alignment in disks}

% RAT paradigm... internal alignment means... external alignment means...
% observations bring the challenge of internal alignment for 1mm grains... 
% could it be wrong alignment?
% NGC1333 IRAS4A also has VLA polarization
% disks with at least two different wavelengths: 

Our results add to the growing picture of disk polarization caused by effectively prolate grains that are aligned toroidally. Currently, we do not have a natural explanation for why grains in the disk behave in this manner. The current RAT alignment paradigm requires the grains to first achieve ``internal'' alignment and then achieve ``external'' alignment. Internal alignment occurs when the internal dissipation of energy in a grain aligns the axis of the largest moment of inertia to the grain's angular momentum direction through Barnett relaxation \citep{Purcell1979ApJ...231..404P}, nuclear relaxation \citep{Lazarian1999ApJ...520L..67L}, or inelastic relaxation \citep[e.g.][]{Purcell1979ApJ...231..404P, Lazarian1999MNRAS.303..673L, Hoang2009ApJ...697.1316H, Hoang2022AJ....164..248H}. External alignment refers to the alignment of the angular momentum to a particular direction in space, like the magnetic field \citep[e.g.][]{Draine1996ApJ...470..551D, Draine1997ApJ...480..633D, Lazarian2007MNRAS.378..910L}, radiation field \citep[e.g.][]{Lazarian2007MNRAS.378..910L}, or gas flow \citep[e.g][]{Lazarian2007ApJ...669L..77L, Reissl2023A&A...674A..47R}. 

The requirement that the polarization-producing grains are effectively prolate indicates that such grains are not internally aligned with their spin axes along the axis of largest moment of inertia; otherwise, the spin would make them effectively oblate when ensemble-averaged. The lack of internal alignment may not be too surprising, especially since large grains of more than $\sim 10$~$\mu$m (in typical densities of protoplanetary disks) have slow internal relaxation that is much longer than the gas randomization timescale making internal alignment difficult \citep{Hoang2009ApJ...697.1316H}. 

Why the effectively prolate grains align with their long axes azimuthally remains a mystery. The Gold mechanism is particularly interesting since the grains should have their long axes aligned to the direction of the dust drift with respect to the gas \citep[e.g.][]{Gold1952MNRAS.112..215G, Lazarian1994MNRAS.268..713L}. The inferred azimuthal alignment direction would suggest dust drift in the azimuthal direction for HL Tau \citep{Yang2019MNRAS.483.2371Y}. However, the Gold mechanism requires supersonic speeds which is not applicable for a protoplanetary disk \citep{Purcell1979ApJ...231..404P, Takeuchi2002ApJ...581.1344T}. 

\subsection{What can produce the near-far side asymmetry?} \label{sec:origin_of_nearfar_side_asymmetry}

Section~\ref{sec:near_far_side_asymmetry} explored the near-far side asymmetry seen in HL Tau. As mentioned in Section~\ref{sec:principal_frame_view}, the far side of the disk is independently determined from the outflow direction. Thus, we discuss if the near-far side asymmetry matches the expected asymmetry from dust scattering of an axisymmetric disk that is inclined, optically thick, and vertically thick \citep{Yang2017MNRAS.472..373Y, Lin2023MNRAS.520.1210L}. The model predicts that along the disk minor axis, $P$ and $p$ should both be stronger along the near side than the far side with the polarization direction parallel to the disk minor axis (i.e, positive $Q'$ and $q'$ following our notation) in the optically thick regions. In translucent regions (optical depth of order unity; such as in the outer disk regions), the polarization direction changes by 90$^{\circ}$ to make $Q'$ negative, but the near side $Q'$ remains more positive than the far side. The asymmetry disappears at larger radii in the optically thin limit. The polarization asymmetry from the data (except the inner regions of Band~4) appears to match the model expectation overall. The $\sim 0.3\%$ difference of $q'$ (Fig.~\ref{fig:nearfar_sides}, 5th column) is also comparable to the model predicted $\sim 0.5\%$ depending on the vertical thickness of the disk \citep{Lin2023MNRAS.520.1210L}.

For Stokes~$I$, the model predicts that the asymmetry should be stronger along the far side than the near side and disappear as the disk becomes optically thin at the outer radii. This is mostly the case for Bands~3 to 7 within 50~au as seen from the negative $f$ (Fig.~\ref{fig:nearfar_sides}, 4th column). The $f \sim -5\%$ for the better resolved Bands~4 to 7 also appears similar to the model predictions depending on the vertical thickness of the disk \citep{Lin2023MNRAS.520.1210L}. However, between 50 to 100~au, the Stokes~$I$ becomes stronger in the near side as seen from the positive $f$ which is not expected from the model. 

One possibility to produce a near side with stronger Stokes~$I$ is if the scattering grains have strong forward scattering \citep{Tazaki2019ApJ...885...52T, Lin2023MNRAS.520.1210L}. Strong forward scattering of grains occurs when the grain size is comparable to or larger than the wavelength of the scattering light. In optically thin regions at outer radii, where much of the radiation travels outwards, scattering of a grain from the near side is more forward scattered, while the photons streaming radially outward in the far side are more backward scattered to reach the observer. In optically thick regions at smaller radii, the effects of forward scattering disappear and the disk retains the near-far side asymmetry in the original case. Thus, $f$ is negative at inner radii, but becomes positive at outer radii. The difference can be $\sim 5$ to $10\%$ depending on the vertical thickness and strength of forward scattering. The observed Stokes~$I$ asymmetry (Fig.~\ref{fig:nearfar_sides}, 4th column) appears to match the model prediction qualitatively and quantitatively. However, strong forward scattering cannot explain the asymmetry of $q'$.

We are thus faced with a conundrum. Scattering without strong forward scattering can produce the asymmetry of $q'$ but not that of Stokes~$I$, while scattering with forward scattering produces the asymmetry of Stokes~$I$ but not that of $q'$. Another puzzle is that the near far side asymmetry for both Stokes~$I$ and $q'$ should decrease as the optical depth decreases with increasing wavelength as the disk becomes optically thinner. While the asymmetry of Stokes~$I$ at Band~3 (Fig.~\ref{fig:nearfar_sides}d) does appear smaller than those at shorter wavelengths, Band~3 also has a much larger beam making it unclear if the difference is truly from optical depth effects. The asymmetry in $q'$, on the other hand, appears to be fairly consistent across wavelengths. We also note the caveat that the center is defined by fitting the disk with a 2D Gaussian. While the asymmetry of $q'$ is not impacted by uncertainties from the center, the asymmetry of Stokes~$I$ as measured by Eq.~(\ref{eq:stokes_I_near_far_asymmetry}) can depend on the center, but precise determination of the center is beyond the scope of this paper.

Other possibilities include substructures or intrinsically non-axisymmetric features either in the surface density distribution and/or in the grain properties. The high angular resolution (sub)mm-continuum images of HL Tau have revealed intricate rings and gaps with radially varying optical depth \citep{Pinte2016ApJ...816...25P, CarrascoGonzalez2016ApJ...821L..16C, CarrascoGonzalez2019ApJ...883...71C}, which is different from the smooth disk models considered thus far and important since the near-far side asymmetries rely on optical depth effects. Furthermore, \cite{ALMApartnership2015} showed that the rings are not concentric which cannot be explained with an inclined axisymmetric disk. How these non-axisymmetric structures manifest as asymmetries in the lower-resolution images (without resolving the rings and gaps) is unclear. Future high angular resolution images across multiwavelengths will be better suited to address these questions. For example, a recent deep, high angular resolution image at Band~7 resolved the polarization from rings and gaps and found additional asymmetries (Stephens et al., in press).

\section{Conclusions} \label{sec:conclusions}

We present and analyze multiwavelength polarization observations of the HL Tau disk at Bands 3, 4, 5, 6, and 7 from ALMA and Q-Band from VLA consolidating HL Tau's position as the protoplanetary disk with the most complete wavelength coverage in resolved dust polarization. Our main results are summarized as follows: 

\begin{enumerate}
    \item New polarization observations using ALMA detected well-resolved polarization at Bands 4, 5, and 7 with angular resolutions of $\sim 0.20\arcsec$, $0.17\arcsec$, and $0.16\arcsec$, respectively. The new VLA Q Band image has a resolution of $\sim 0.15\arcsec$ and marginally detects a few polarization vectors. The new data strengthens the case for a smooth systematic transition from unidirectional polarization direction to an azimuthal direction as the wavelength increases. 
    
    \item The polarization transition is further evidence of scattering prolate grains aligned toroidally in the disk. We disentangle the polarization from scattering and the elongated grains' thermal emission through the azimuthal variation of polarization from a simple model. The constant component from scattering decreases slowly with increasing wavelength, while the thermal component, which causes azimuthal variation, increases with increasing wavelength. The weak dependence of the scattering spectrum is more consistent with large, porous grains than compact small grains. 
    
    \item The few polarization detections at Q-band are also consistent with toroidally aligned grains by comparing the expected polarization angles. The polarization fraction is higher, at $\sim 7\%$, and suggests that the intrinsic polarization of grains can be $\sim 10\%$ after correcting for projection of the grain. 
    
    \item We find a consistent near-far side asymmetry in the polarization fraction and Stokes~$I$ at ALMA Bands 3, 4, 5, 6, and 7. The near-far side asymmetry of the polarization can be explained by optically thick and geometrically thick disk. However, the near-far side asymmetry in the Stokes~$I$ is harder to explain and deserves further exploration.  
\end{enumerate}

\section*{Acknowledgements}

Z.-Y.D.L. acknowledges support from NASA 80NSSC18K1095, the Jefferson Scholars Foundation, the NRAO ALMA Student Observing Support (SOS) SOSPA8-003, the Achievements Rewards for College Scientists (ARCS) Foundation Washington Chapter, the Virginia Space Grant Consortium (VSGC), and UVA research computing (RIVANNA). Z.-Y.L. is supported in part by NASA 80NSSC20K0533 and NSF AST-2307199 and AST-1910106. C.C.-G. acknowledges support from UNAM DGAPA-PAPIIT grant IG101321 and from CONACyT Ciencia de Frontera project ID 86372. L.W.L. acknowledges support from NSF 1910364 and NSF 2307844. D.S.-C. is supported by an NSF Astronomy and Astrophysics Postdoctoral Fellowship under award AST-2102405. 

This paper makes use of the following ALMA data: ADS/JAO.ALMA\#2016.1.00115.S, ADS/JAO.ALMA\#2019.1.00134.S, ADS/JAO.ALMA\#2019.1.00162.S, ADS/JAO.ALMA\#2019.1.01051.S. ALMA is a partnership of ESO (representing its member states), NSF (USA), and NINS (Japan), together with NRC (Canada), MOST and ASIAA (Taiwan), and KASI (Republic of Korea), in cooperation with the Republic of Chile. The Joint ALMA Observatory is operated by ESO, AUI/NRAO and NAOJ. The National Radio Astronomy Observatory is a facility of the National Science Foundation operated under cooperative agreement by Associated Universities, Inc.

%%%%%%%%%%%%%%%%%%%%%%%%%%%%%%%%%%%%%%%%%%%%%%%%%%
\section*{Data Availability}

Data underlying this article is available from the corresponding author upon request.

%The inclusion of a Data Availability Statement is a requirement for articles published in MNRAS. Data Availability Statements provide a standardised format for readers to understand the availability of data underlying the research results described in the article. The statement may refer to original data generated in the course of the study or to third-party data analysed in the article. The statement should describe and provide means of access, where possible, by linking to the data or providing the required accession numbers for the relevant databases or DOIs.



\bibliographystyle{mnras}
\bibliography{main} % if your bibtex file is called example.bib

%\begin{thebibliography}{99}
%\end{thebibliography}

%%%%%%%%%%%%%%%%%%%%%%%%%%%%%%%%%%%%%%%%%%%%%%%%%%

%%%%%%%%%%%%%%%%% APPENDICES %%%%%%%%%%%%%%%%%%%%%

\appendix

\section{Stokes~$V$ Images} \label{sec:stokes_V}

\begin{figure}
    \centering
    \includegraphics[width=0.72\columnwidth]{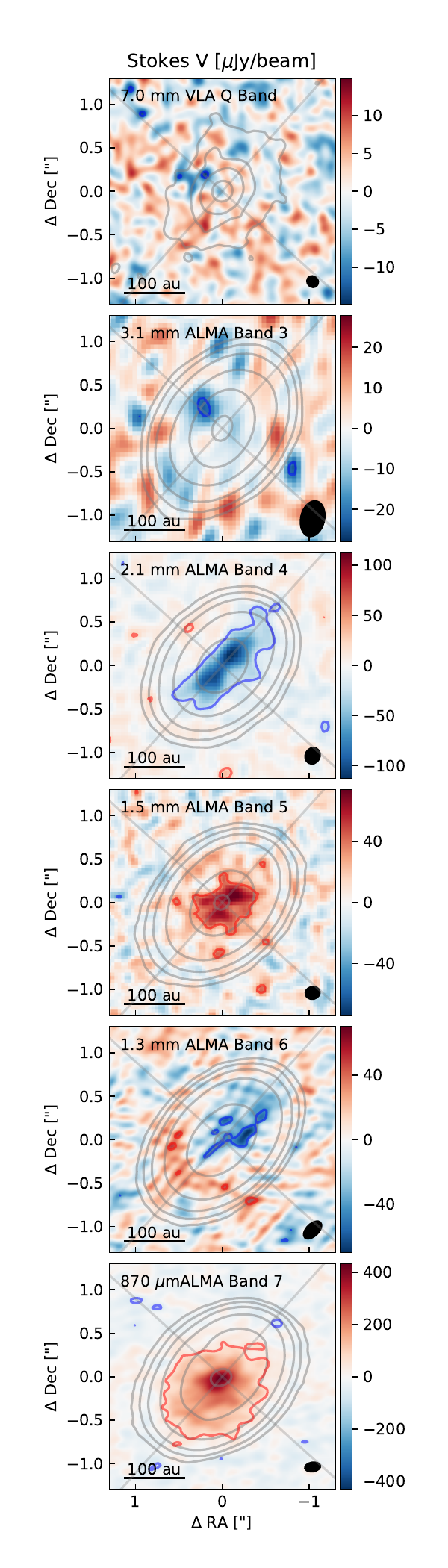}
    \caption{The Stokes~$V$ images plotted in a similar manner as Fig.~\ref{fig:obs_stokes_1}. The color scales are plotted such that the white corresponds to the zero level. The $-3\sigma_{V}$ and $3\sigma_{V}$ levels are marked by blue and red contours, respectively.}
    \label{fig:obs_stokesV}
\end{figure}

The noise level ($\sigma_{V}$) and peak absolute value of Stokes~$V$ are listed in Table~\ref{tab:stokesV}. Fig.~\ref{fig:obs_stokesV} shows the Stokes~$V$ images across each band. Unlike the smooth transitions from wavelength to wavelength for Stokes~$I$, $Q$, and $U$, Stokes~$V$ varies with wavelength more erratically. At $7.1$~mm, a slight negative Stokes~$V$ of $\sim 3 \sigma_{V}$ is detected to the northeast, which is similar to the image at $3.1$~mm. At $2.1$~mm, the $\sim 15 \sigma_{V}$ detection of negative Stokes $V$ appears to have two peaks along the disk major axis. However, at $1.5$~mm, the Stokes~$V$ becomes positive and mostly concentrated at the center with $\sim 6 \sigma$. Another change happens at $1.3$~mm in which case the southeast half of the disk is mostly positive and the northwest half is negative. Finally, at $870 \mu$m, Stokes~$V$ is positive and concentrated at the center with a peak of $\sim 21 \sigma_{V}$. However, the $870 \mu$m image from \cite{Stephens2017} shows a negative Stokes~$V$. ALMA is known to have significant instrumental errors in Stokes~$V$, which is primarily due to beam squint. The inconsistency between the Band~7 images for two different epochs suggests that the current ALMA Stokes $V$ detections are largely due to instrumental effects.

According to the ALMA technical handbook, the minimum detectable degree of circular polarization for ALMA is $1.8\%$ of the peak flux on-axis based on the ALMA technical handbook. Indeed, the ALMA peak $|V|$ detections all fall below the minimum detectable threshold (which are $\sim 457, 495, 850, 1470, 2180$~$\mu$Jy beam$^{-1}$ for Bands 3 to 7, respectively). 
%\red{The minimum detectable degree of circular polarization is 1.8\% of the peak flux on-axis. Imaging is affected by beam-squint and therefore, ALMA advice is to restrict the imaging to the inner 1/10 of the primary beam FWHM. I have estimated the upper limits for a spurious signal at all wavelengths, and they are far larger than the Stokes V peaks in the maps (1.8\% levels for Bands 3 ,4 , 5, 6, and 7: 457, 495, 850, 1472, and 2181 $\mu$Jy, respectively. The larger the Stokes I peak, the larger the detection threshold.)}

\begin{table}
    \centering
    \begin{tabular}{c c c}
        \hline 
        Band & $\sigma_{V}$ & Peak $|V|$ \\
        & $\mu$Jy/beam & $\mu$Jy/beam \\
        (1) & (2) & (3) \\
        \hline
        Q & 4.1 & 15 \\
        3 & 7.0 & 23 \\
        4 & 7.7 & 113 \\
        5 & 12 & 74 \\
        6 & 14 & 70 \\
        7 & 21 & 430 \\
        \hline 
    \end{tabular}
    \caption{The basic image statistics for Stokes~$V$. Column 1: Name of the wavelength band. Column 2: The noise level for Stokes~$V$. Column 3: Peak of the absolute value of Stokes~$V$ image. }
    \label{tab:stokesV}
\end{table}

\section{Derivation for the azimuthal variation of thermal polarization} \label{sec:thermal_polarization_in_principal_frame}
We supply a few more details on the derivation of the thermal polarization shown in Section~\ref{sec:linear_decomposition}. The basis was presented in \cite{Lin2022}, but given the difference in the definition of the Stokes reference frames used in this paper, we provide an explicit derivation for clarity. As mentioned in the main text, the reference frames strictly follow the IEEE definition. 

\begin{figure}
    \centering
    \includegraphics[width=\columnwidth, trim={0.5cm 0.5cm 0.5cm 0.5cm},clip]{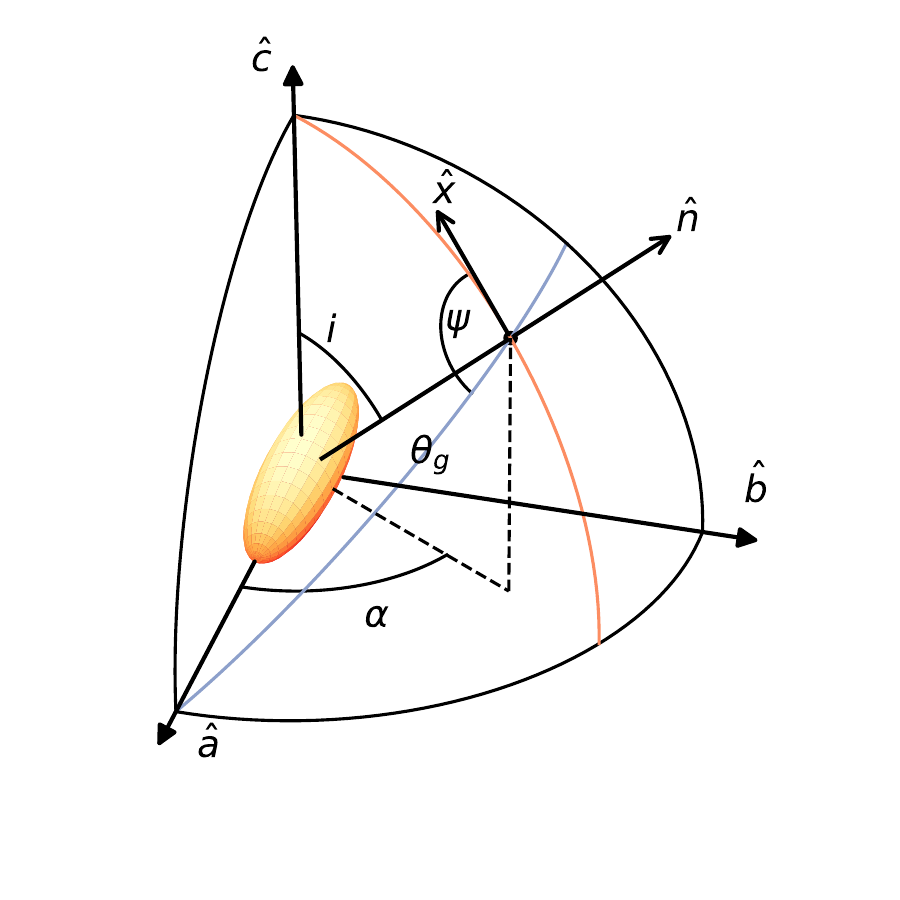}
    \caption{
        Schematic of the relation between a prolate grain to the observer. The $\hat{a}$, $\hat{b}$, and $\hat{c}$ unit vectors form the coordinates centered on a grain represented by a prolate. $\hat{a}$ is parallel to the axis of symmetry of the grain. $\hat{c}$ is parallel to the $Z$-axis of the disk. $i$ is the inclination. The orange arc is the meridian passing through $\hat{n}$ and $\hat{c}$. The blue arc passes through $\hat{a}$ and $\hat{n}$. The two planes form an angle $\psi$. $\hat{x}$ is parallel to the $x$-axis of the principal frame. 
    }
    \label{fig:grain_to_sky}
\end{figure}

Fig.~\ref{fig:sky_schematic} in the main text showed the relation between the principal frame and the disk with aligned grains. Fig.~\ref{fig:grain_to_sky} shows the relation between the principal frame and each grain located at a different location in the disk. Consider the coordinates around a single grain with unit vectors $\hat{a}$, $\hat{b}$, and $\hat{c}$, where $\hat{a}$ is along the axis of symmetry of the grain and $\hat{c}$ is parallel to the $Z$-axis of the disk. The inclination, $i$, is simply the angle between $\hat{c}$ and the direction to the observer, $\hat{n}$. Following Fig.~\ref{fig:sky_schematic}, one can easily see that $\hat{x}$ is in the direction of the $x$-axis of the principal frame. The $y$-direction of the image frame is not shown in the plot to avoid clutter, but it is in the direction of $\hat{n} \times \hat{x}$. The azimuthal angle in this coordinate, $\alpha$, is the angle between $\hat{a}$ and the projection of $\hat{n}$ onto the $\hat{a}$-$\hat{b}$ plane.

Depending on the location along the disk azimuth $\Phi$, the grain can be seen edge-on or closer to pole-on, which gives the azimuthal variation of $p$ seen in the image (Sec.~\ref{sec:linear_decomposition}). We use $\theta_{g}$ to denote the viewing angle of the grain, which is the angle from $\hat{a}$ to $\hat{n}$. Since the prolate grains are assumed to be toroidally aligned, one can derive that
\begin{align}
    \cos \theta_{g} = \hat{n} \cdot \hat{\Phi} = \sin i \sin \Phi .
\end{align}

The thermal polarization fraction for a grain, $t_{p}$, from Eq.~(\ref{eq:thermal_polarization_approximation}) simply gives the magnitude of $p$ given some $\theta_{g}$. To obtain the $q$ and $u$, one needs to define a reference frame. We can start by defining a Stokes reference frame (which we call the ``grain frame") in the same $\hat{x}$-$\hat{y}$ plane, but with rotated such that the new $\hat{x}_{g}$ is in the plane formed by $\hat{a}$ and $\hat{n}$. The angle between $\hat{x}$ and $\hat{x}_{g}$ is $\psi$. The Stokes parameter between the grain frame and the principal frame only requires a rotation of the Stokes parameters and we have 
\begin{align} \label{eq:rotate_thermal_polarization_in_principal_frame}
    q'_{t} &= t_{p} \cos 2 \psi \\
    u'_{t} &= t_{p} \sin 2 \psi. 
\end{align}
We can express $\psi$ from geometrical arguments and obtain
\begin{align}
    \cos \psi &= - \frac{ \cos \theta_{g} \cos i }{ \sin \theta_{g} \sin i } \label{eq:cos_psi} \\
    \sin \psi &= \frac{ \sin \alpha }{ \sin \theta_{g} } \label{eq:sin_psi} 
\end{align}
\citep[see][]{Lin2022}.  
Since the grain is toroidally aligned (i.e., $\hat{a} = \hat{\Phi}$) and given $\hat{n}$ defined in Fig.~\ref{fig:sky_schematic}, one can find that $\alpha=\pi / 2 - \Phi$. Using Eq.~(\ref{eq:cos_psi}), (\ref{eq:sin_psi}), we get a fairly simple expression of $t_{p}$ in the principal frame: 
\begin{align} 
    q'_{t} &= t (\cos^{2} i \sin^{2} \Phi - \cos^{2} \Phi) \label{eq:thermal_q_principal_frame} \\
    u'_{t} &= - t \cos i \sin 2 \Phi \label{eq:thermal_u_principal_frame}. 
\end{align}
These are the contributions from thermal polarization to Eq.~(\ref{eq:azimuthal_q_linear_decomposition}) and (\ref{eq:azimuthal_u_linear_decomposition}) in the main text. 

\section{Posterior Probability Distribution} \label{sec:corner_plots}

\begin{figure*}
     \centering
     \begin{subfigure}[b]{0.33\textwidth}
         \centering
         \includegraphics[width=\textwidth]{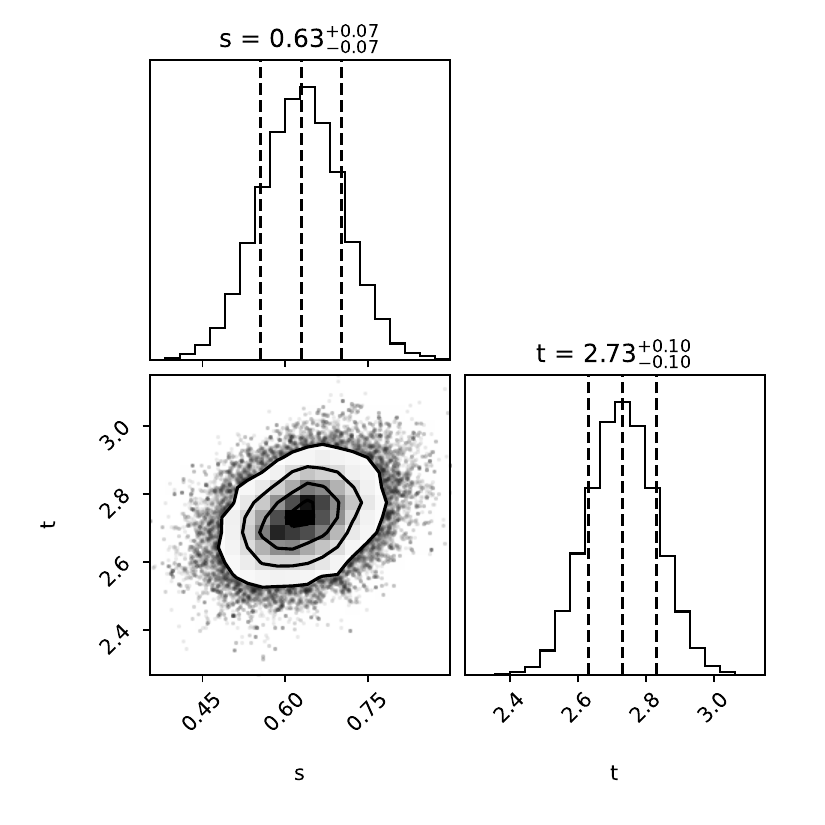}
         \caption{Band 3}
         \label{fig:corner_b3}
     \end{subfigure}
     \hfill
     \begin{subfigure}[b]{0.33\textwidth}
         \centering
         \includegraphics[width=\textwidth]{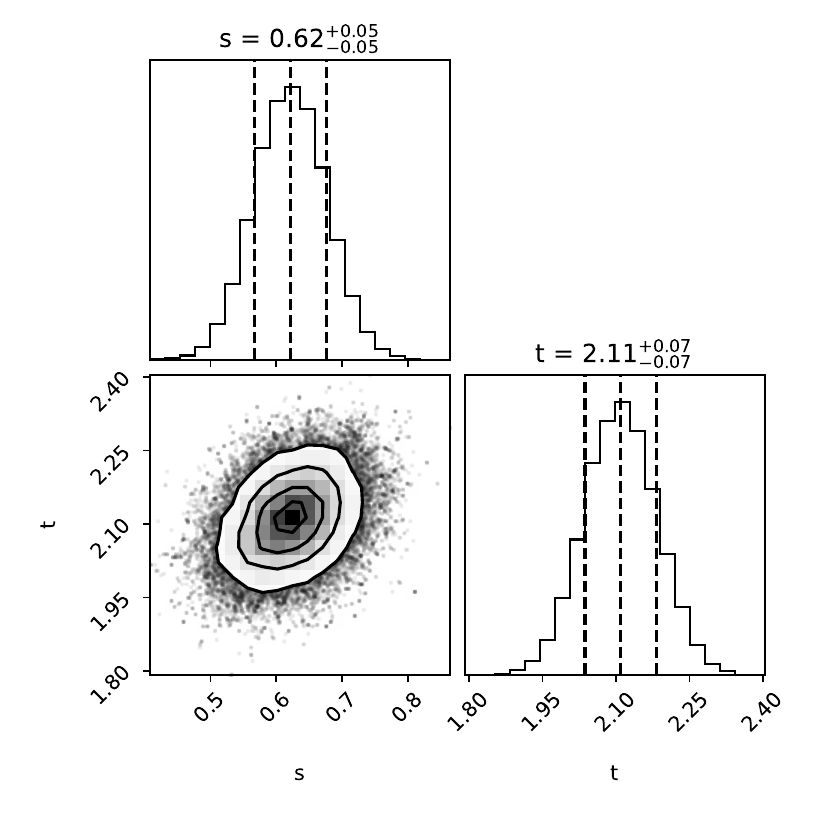}
         \caption{Band 4}
         \label{fig:corner_b4}
     \end{subfigure}
     \hfill
     \begin{subfigure}[b]{0.33\textwidth}
         \centering
         \includegraphics[width=\textwidth]{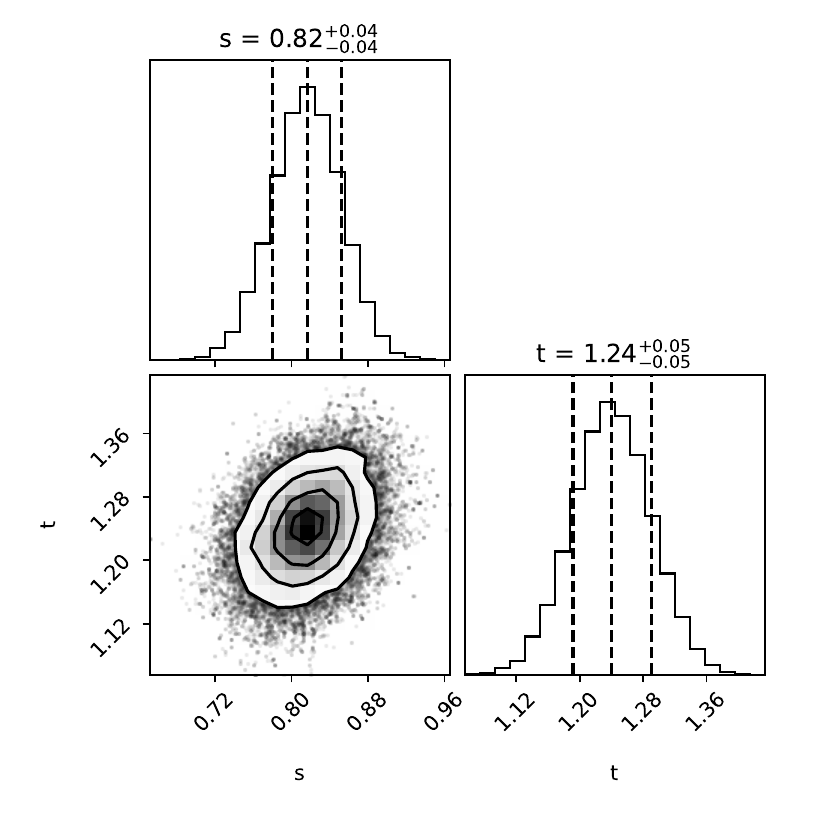}
         \caption{Band 5}
         \label{fig:corner_b5}
     \end{subfigure}
     \\[\smallskipamount]
          \begin{subfigure}[b]{0.33\textwidth}
         \centering
         \includegraphics[width=\textwidth]{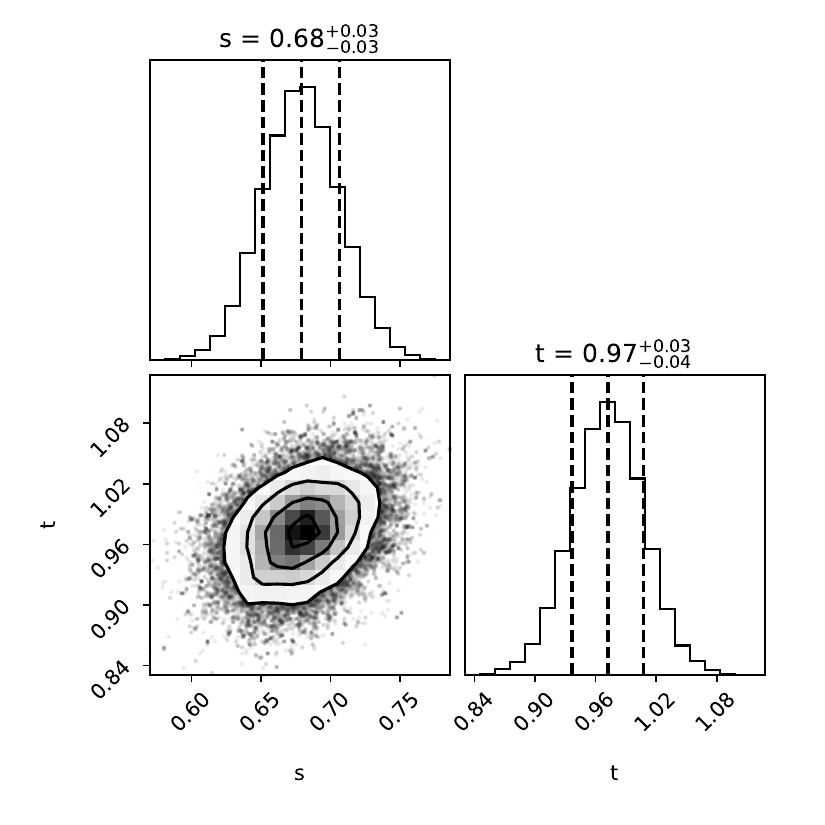}
         \caption{Band 6}
         \label{fig:corner_b6}
     \end{subfigure}
     \hfill
     \begin{subfigure}[b]{0.33\textwidth}
         \centering
         \includegraphics[width=\textwidth]{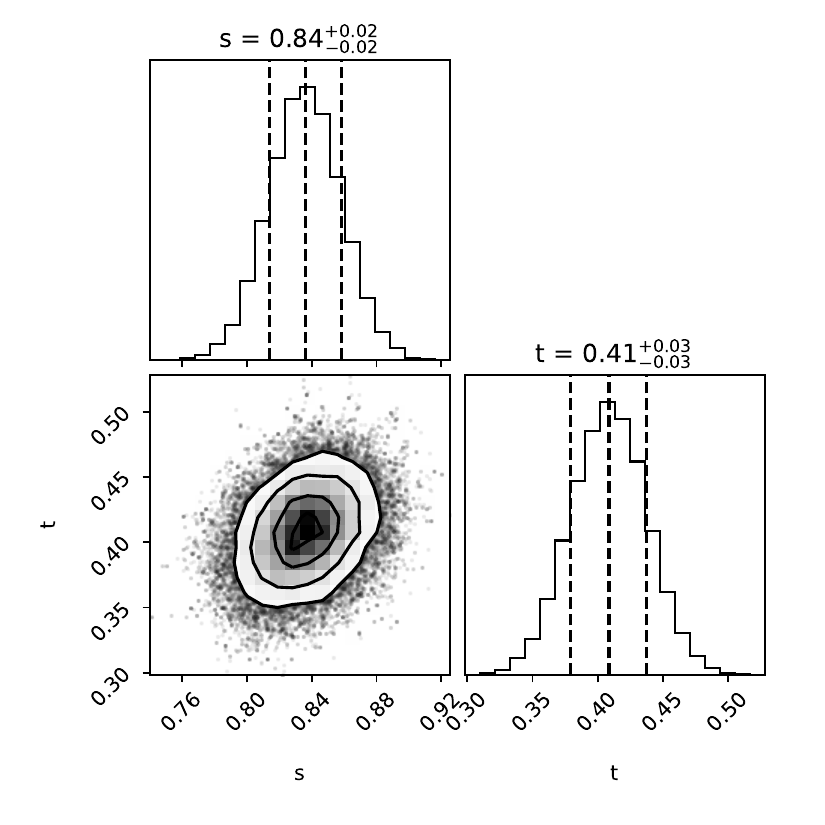}
         \caption{Band 7}
         \label{fig:corner_b7}
     \end{subfigure}
     \hfill
     \begin{subfigure}[b]{0.33\textwidth}
         \centering
         \includegraphics[width=\textwidth]{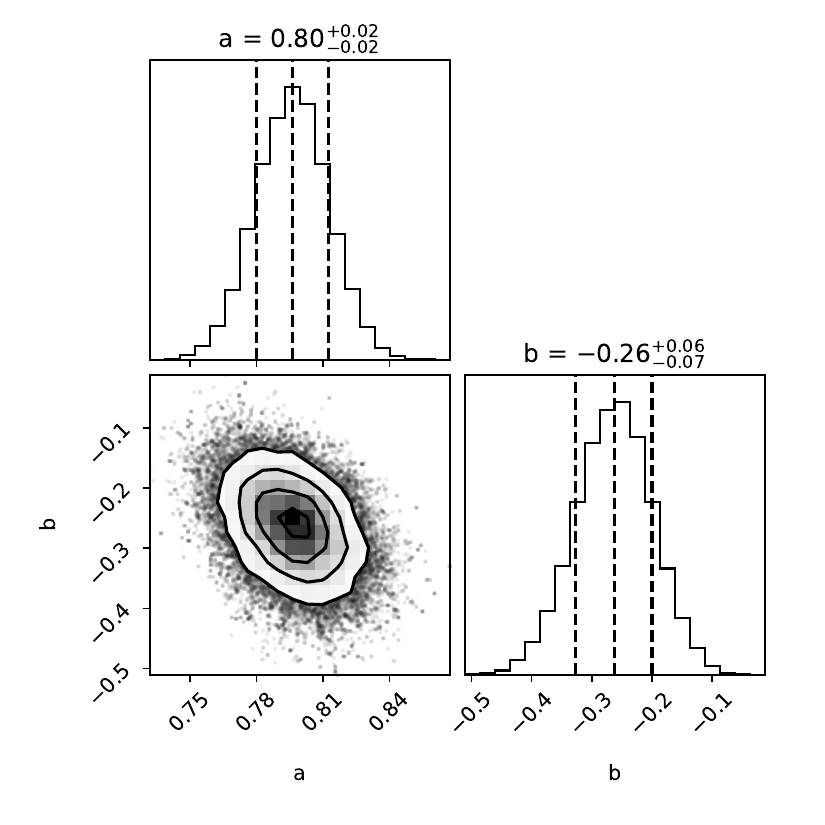}
         \caption{Parameters for the $s$-spectrum.}
         \label{fig:corner_s_spectrum}
     \end{subfigure}
    \caption{One- and two-dimensional posterior probability distribution from \textit{emcee}. Panels a to e: Results from fitting the azimuthal profile of $q'$ and $u'$ from bands 3 to 7, respectively. Panel f: Results from fitting the $s$-spectrum (corresponding to Fig.~\ref{fig:st_spectrum}a and its discussion). }
    \label{fig:corner_plots}
\end{figure*}

In Section~\ref{sec:linear_decomposition}, we fit the azimuthal profile of $q'$ and $u'$ with the linear decomposition model using \textit{emcee}. Fig.~\ref{fig:corner_plots}a through e show the resulting one- and two-dimensional posterior probability distribution at each wavelength. Fig.~\ref{fig:corner_plots}f is the result of fitting the $s$-spectrum.

%%%%%%%%%%%%%%%%%%%%%%%%%%%%%%%%%%%%%%%%%%%%%%%%%%

% Don't change these lines
\bsp	% typesetting comment
\label{lastpage}
\end{document}